\documentclass[conference]{IEEEtran}
\usepackage{times}

\usepackage{url}
\usepackage[utf8]{inputenc} 
\usepackage[T1]{fontenc}    
\usepackage{url}            
\usepackage{booktabs}       
\usepackage{amsfonts}       
\usepackage{nicefrac}       
\usepackage{microtype}      
\usepackage{xcolor}
\newcommand{\blue}[1]{\textcolor{blue}{#1}}
\newcommand{\red}[1]{\textcolor{red}{#1}}

\usepackage{multirow}
\usepackage[most]{tcolorbox}
\usepackage{pifont}
\newcommand{\cmark}{\ding{51}}%
\newcommand{\xmark}{\ding{55}}%
\usepackage{wrapfig}
\usepackage{amsmath}
\usepackage{adjustbox}
\usepackage{lineno}
\usepackage{rotating}
\newcommand{\dataset}{Social Human Robot Embodied Conversation
(SHREC) Dataset}
\newcommand{\datasets}{Social Human Robot Embodied Conversation
(SHREC) Dataset }

\newcommand{\datasetshorts}{SHREC Dataset }

\usepackage[numbers]{natbib}
\usepackage{multicol}
\usepackage[bookmarks=true]{hyperref}
\usepackage{amsmath}
\usepackage{graphicx}

\usepackage[numbers]{natbib}
\usepackage{multicol}
\usepackage[bookmarks=true]{hyperref}

\begin{document}

\title{Social Human Robot Embodied Conversation (SHREC)   Dataset: Benchmarking Foundational Models' Social Reasoning}

\author{Author Names Omitted for Anonymous Review. Paper-ID 418}



%
\author{\parbox{0.9\linewidth}{
\centering{Dong Won Lee$^{\diamondsuit}$~~~Yubin Kim$^{\diamondsuit}$~~~  Sooyeon Jeong$^{\heartsuit}$~~~  Denison Guvenoz$^{\heartsuit}$~~~  \\
Parker Malachowsky$^{\diamondsuit}$~~~ Louis-Philippe Morency$^{\spadesuit}$~~~ \\
Cynthia Breazeal$^{\diamondsuit}$~~~  Hae Won Park$^{\diamondsuit}$}\\
\centering{\rm $^\heartsuit$Purdue University
~~$^\spadesuit$Carnegie Mellon University \\ 
~~$^\diamondsuit$Massachusetts Institute of Technology} \\
\vspace{0.5em}
{\textbf{Dataset:}~\href{https://huggingface.co/collections/MIT-personal-robots/shrec-6822cd769bca5b8ed9ae299f}{\texttt{https://huggingface.co/collections/MIT-personal-robots/shrec}}} \\
{\textbf{Code: }~\href{https://github.com/mitmedialab/SHREC}{\texttt{https://github.com/mitmedialab/SHREC}}}
}
}

\maketitle

\begin{abstract}
Our work focuses on the social reasoning capabilities of foundational models for real-world human–robot interactions. We introduce the \dataset, a benchmark of $\sim$400 real-world human-robot interaction videos and over 10,000 annotations, capturing robot social errors, competencies, underlying rationales, and corrections. Unlike prior datasets focused on human–human interactions, the \datasetshorts uniquely highlights the social  challenges faced by real-world social robots  such as emotion understanding, intention tracking, and conversational mechanics. Moreover, current foundational models struggle to recognize these deficits, which manifest as subtle, socially situated failures. To evaluate AI models’ capacity for social reasoning, we define eight benchmark tasks targeting critical areas such as (1) detection of social errors and competencies, (2) identification of underlying social attributes, (3) comprehension of interaction flow, and (4) providing rationale and alternative correct actions. Experiments with state-of-the-art foundational models, alongside human evaluations, reveal substantial performance gaps—underscoring the difficulty and providing directions in developing socially intelligent AI.
\end{abstract}

\section{Introduction}
\label{sec:intro}

\begin{figure}
    \begin{center}
    \includegraphics[width=\linewidth]{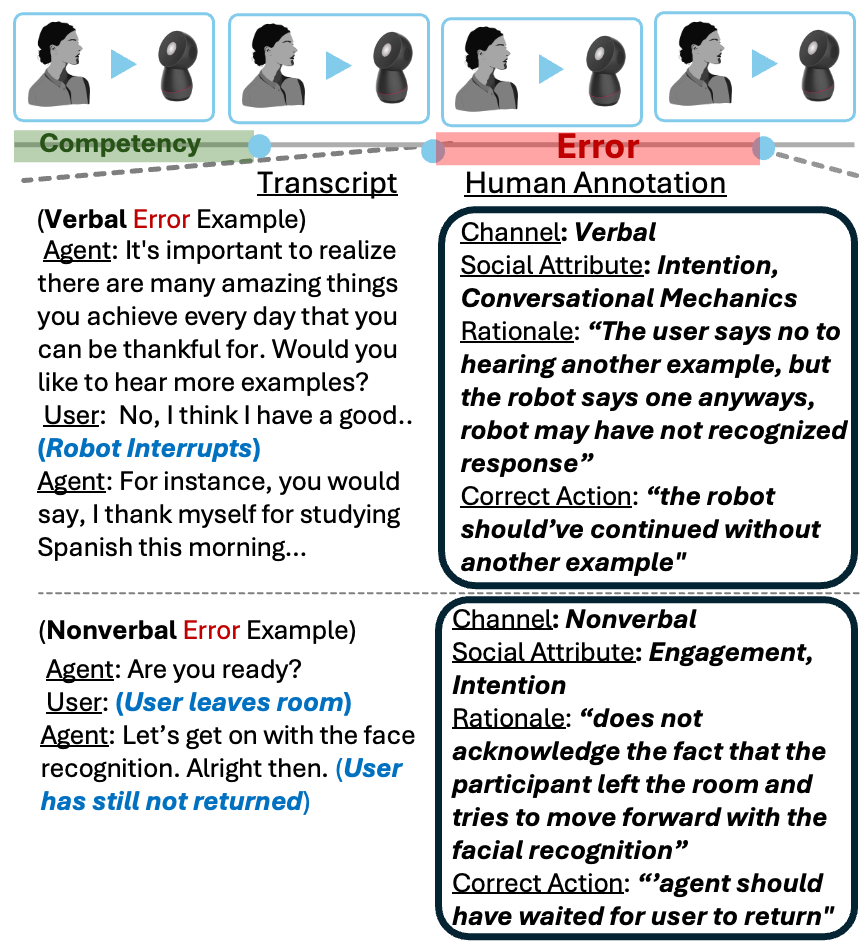}
    \vspace{-7mm}
    \caption{\datasetshorts dataset offers real-world \textbf{Social Human Robot  Embodied Conversation} videos and annotations of errors and competencies, the channel and type of social attribute, along with rationale and possible corrective actions. (Top) Error sourced from verbal (audio) channel, (Bottom) Error sourced from non-verbal (visual) channel.}
    \label{fig:first_fig}
    \end{center}
    \vspace{-8mm}
\end{figure}

In this work, we aim to advance the social reasoning capabilities of physically embodied AI agents, particularly tabletop social robots, engaged in real-world, socially interactive conversations. To this end, we focus on understanding and modeling both social competencies (i.e. desirable behaviors) and social errors (i.e. norm violations or failures in interaction) that arise during natural human-robot interactions. While prior work in social intelligence has primarily centered on human–human interaction datasets \citep{mathur2025social, zadeh2019social, wilf2023social}, these settings do not capture the unique challenges that arise when robots interact with humans. Unlike humans, embodied AI agents may have varying and lacking  socio-cognitive skills such as emotion understanding, belief tracking, or conversational coordination \citep{lake2017building, deng2023socratis, bhattacharyya-wang-2025-evaluating, arora2025talking}, leading to failure cases that are subtle, socially situated, and poorly represented in existing resources.

To this end, we introduce the \textbf{\dataset}, a dataset of 400+ videos capturing natural conversational interactions between a human and a physical tabletop social robot, making it, to the best of our knowledge, one of the largest real-world human–social robot interaction benchmark dataset available to date (see App. Tab. \ref{tab:compare}). 
We focus on a tabletop social robot, which is a widely used class of physically embodied agents, as a principled starting point for studying real-world social reasoning, and introduce a benchmark dataset that addresses the scarcity of real-world social robot interaction data available for advancing research in this area. The dataset is accompanied by 10K+ human annotations of the robot's social errors (undesirable behaviors) competencies (desirable behaviors), rationale and corrective actions. These annotations are grounded from  prior HRI taxonomies \citep{tian2021taxonomy,fitrianie2022artificial} of seven key social attributes that underpin social interactions, emotion, engagement, conversational mechanics, knowledge state, intention, social relationships, and norms.

The \datasetshorts offers a novel resource to assess foundational models' capabilties in identifying a social robot's social errors and competencies in real-world human-robot  interactions, filling a gap not addressed by current social intelligence benchmarks. To systematically evaluate social reasoning of state-of-the-art AI models, we propose eight benchmark tasks, spanning four core dimensions: (1) errors and competency detection in the robot's behavior, (2) social attribute identification related to the errors and competencies, (3) interaction progression  reasoning, and (4) rationale and correction reasoning (outlined in Sec. \ref{sec:tasks}). Beyond assessing models’ social reasoning, generalization, and robustness in real-world, multimodal, and socially grounded settings \citep{zhou2024sotopiainteractiveevaluationsocial,davis2015commonsense,lake2017building,sap2019atomic,ross2022does,ludan2023explanation}, these tasks also serve as structured probes for fundamental representation learning challenges. They require multimodal alignment to integrate visual, auditory, and linguistic signals into representations for social understanding; reasoning to infer how specific behaviors influence downstream interaction trajectories; and reward-learning to align embodied agents to human social preferences. Our benchmark provides not only a principled testbed for advancing socially intelligent foundation models but also a new resource investigate these core learning problems.

We benchmarked 18 state-of-the-art LLMs and VLMs, including ChatGPT \citep{hurst2024gpt} and Gemini \citep{team2024gemini} variants, on these tasks. While some models excel in specific subtasks, none perform  well across the board. Notably, the gap between model and human performance remains substantial, underscoring the challenge and novelty of our benchmark. These findings highlight \datasets as a valuable testbed for diagnosing and improving the social reasoning abilities of embodied AI agents, and for guiding the development of reward models and evaluators aligned with social intelligence \citep{ouyang2022training, zhou2024sotopiainteractiveevaluationsocial, lee2023rlaif, chen2024self, zheng2023judging}.

\section{Related Work}
\subsection{Datasets for Social Interaction Analysis} Several datasets have been developed to analyze human social interactions, many of which focus on conversational data or multimodal behaviors. For example, the MELD (Friends TV Series) Dataset \citep{poria-etal-2019-meld} and the CMU Multimodal Opinion Sentiment and Emotion Intensity (CMU-MOSEI) dataset \citep{zadeh2018multimodal} provide annotated multimodal data for studying emotions and sentiment in dialogue. More closely related to our work, the SocialIQA \citep{sap2019socialiqa} dataset introduced 38,000 multiple-choice questions derived from the ATOMIC \citep{sap2019atomic} knowledge graph, which is a large-scale graph of commonsense knowledge. These questions are aimed at testing models' understanding of social norms, intentions, and emotional responses in social scenarios. Furthermore, the Social-IQ Dataset \citep{zadeh2019social} was designed to evaluate social intelligence in AI with human-to-human interaction videos, including multimodal question-answer pairs that assess the ability to understand and respond to social situations effectively. A recent benchmark, SOCIAL GENOME \citep{mathur2025social}, evaluates multimodal models’ ability to generate grounded social reasoning traces from videos, incorporating fine-grained cues and external knowledge. Our benchmark includes around 400 videos, consisting of 3600 minutes of real-world human–robot interaction footage, this scale is on par with or larger than many multimodal datasets in social interaction research (e.g., Social Genome \cite{mathur2025social}: 280 minutes, Social-IQ \cite{zadeh2019socialiq}: 1200 minutes, MEmoR \cite{shen2020memor}: 2800 minutes, CMU-MOSEI \cite{zadeh2018cmu}: 3900 minutes). We also refer the reviewer to Appendix D, where we provide a comparative table of related works for clarity. Importantly, what distinguishes our dataset is that it is collected from real-world deployments of physically embodied social robots, offering naturalistic, longitudinal interactions, an extremely scarce setting, with, to the best of our knowledge, no publicly available data at this scale with equally comprehensive annotations.

\subsection{Frameworks for Annotating Social Errors in HRI} Previous work in Human-Robot Interaction (HRI) has defined social errors as instances where atypical robot behavior results in violations of social norms \citep{giuliani2015systematic}. These studies primarily focused on nonverbal cues, annotating both the robot’s behavior and the human social signals, such as speech, head movements, gestures, facial expressions, and body posture, during error episodes. A taxonomy to systematically categorize social errors and competencies in HRI was developed, which forms the foundation for our own approach \cite{tian2021taxonomy}. Similarly, the ASA questionnaire was proposed to assess social intelligence in artificial agents, leveraging human feedback collected across interaction trials \cite{fitrianie2022artificial}. While these efforts have proposed valuable conceptual frameworks, our work is the first to contribute a large-scale, aligned dataset of real-world HRI episodes annotated for social errors and competencies.

\subsection{Social Interaction Analysis with Language Models} Several previous studies have focused on the social reasoning capabilities of language-model social agents. Theory of Mind (ToM) has been tested through a variety of tasks, often times through measuring an LLM's ability to understand others' mental states using a series of reasoning-specific tasks \citep{ullman2023large}, or identifying social errors and understanding the perspectives of participants through faux pas tasks \citep{shapira2023well}. Frameworks such as COKE utilize contextual meanings of input entities to more accurately map knowledge graphs to be used in LLMs \citep{wu2023coke}, and the SOCIALIQA benchmark similarly provides a collection of commonsense questions with appropriate and inappropriate responses to be used in LLMs \citep{sap2019socialiqa}. In studying emotional understanding, psychometric assessments have been developed to measure emotional understanding of an LLM based on a given scenario \citep{wang2023emotional}. Benchmarks such as Socratis measured emotional intelligence utilizing a repository of emotional reactions and appropriate scenarios \citep{deng2023socratis}.


\section{SHREC: Dataset of Human Robot Social Embodied Conversation} 
\label{sec:taxonomy}

Our dataset consists of 10,353 annotations from 403 interaction videos spanning over 3,500 minutes. Under a newly accepted IRB protocol, we annotated and anonymized data on three prior human-robot interaction studies \citep{shen2024empathicstories++, jeong2023deploying, jeong2020robotic} to be shared for dissemination. To enable public access while preserving participant privacy, we follow institutional IRB procedures and release the dataset under gated access. All personally identifiable information (PII) in transcripts is filtered. For video anonymization, we leverage FRESCO \citep{yang2024fresco}, a zero-shot video-to-video diffusion framework, to perform stylized face transfer. FRESCO’s spatial-temporal consistency allows us to reliably replace participant faces while maintaining coherence across frames. This ensures that the social signals (e.g., gaze, affect) critical for interaction analysis are preserved while protecting individual identities through high-fidelity anonymization. In selecting FRESCO, we compared several anonymization techniques on the same interaction clips, generating matched variants from the raw videos and evaluating them with the identical pipeline (see Appendix~\ref{app:anon_exp}). Most vision-language models (VLMs) operate at 1 Hz, which is the default temporal resolution used in our benchmark. Hence, we release video frames sampled at 1 Hz to align with common VLM processing rates and ensure compatibility across models. \footnote{To support community interest in fine-grained temporal analysis, we will release an extended addendum with 15Hz videos (processing is underway and sample videos are included in the supplementary). We maintain a private hold-out set of $\sim$40 videos to prevent data contamination and enable future benchmarking and evaluations.}

We describe the three real-world studies that form the foundation of the SHREC dataset, all of which used the \textbf{Jibo social robot} \cite{park2024jibo} as the embodied interaction platform. Jibo is a tabletop social robot originally designed as a daily companion for the home. Its embodiment supports speech-based interaction, expressive gaze, animated idle behaviors such as blinking and looking around, and attention behaviors that orient the robot toward sound sources or people’s faces. These capabilities enable Jibo to proactively initiate lightweight social interactions, offer greetings or activities, and maintain a sense of social presence during interaction. At the same time, Jibo’s embodiment imposes important constraints: it is non-mobile, has no arms or object-manipulation capabilities, and expresses social behavior primarily through speech, gaze orientation, screen-based display, and body/head motion. Firstly, \textbf{Empathic++} \citep{shen2024empathicstories++}: a ChatGPT-powered social robotic agent acts as an empathic companion, facilitating the exchange of emotionally meaningful stories using narrative therapy techniques. The goal is to enhance users’ feelings of connection and belonging through emotionally attuned interaction. \textbf{Wellness-Dorm} \citep{jeong2020robotic}: A socially assistive robot was deployed as a positive psychology coach for college students living in on-campus dormitories. The robot was manually scripted with seven intervention types grounded in established principles of positive psychology, such as gratitude, strengths-based reflection, and goal setting. \textbf{Wellness-Home} \citep{jeong2023deploying}: Positive psychology robots were deployed in participants’ homes. All three datasets are ready for dissemination (We kindly  recommend the reader to view the supplementary materials for examples of the dataset and demographic information).

\subsection{Human Annotations}
As shown in Fig. \ref{fig:first_fig} and App. \ref{app:instructions}, annotators watch videos of conversational human-robot interactions and are asked to  detect segments which manifest a \textbf{\emph{social error or competency}}. \textbf{\emph{Social Competence}} is defined as the behaviors where the agent successfully conduct social interactions by being aware of and identifying social-emotional cues, processing such cues, and expressing a user-expected response to these cues \citep{halberstadt2001affective}. \textbf{\emph{Social Error}} is defined as the behaviors where the robot deviates from the desired behaviors expected by a user and degrades the user's perception of a robot's social competence \citep{tian2021taxonomy}. Then, they are asked to identify which \textbf{\emph{social attribute}} the segment is related to (described in Sec. \ref{ssec:attributes}) and offer a  \textbf{\emph{rationale}} of why they believe so. If the segment was an error, they are asked to suggest an alternative \textbf{\emph{correct action}}.


\begin{figure*}[t]
    \begin{center}
    \includegraphics[width=1\textwidth]{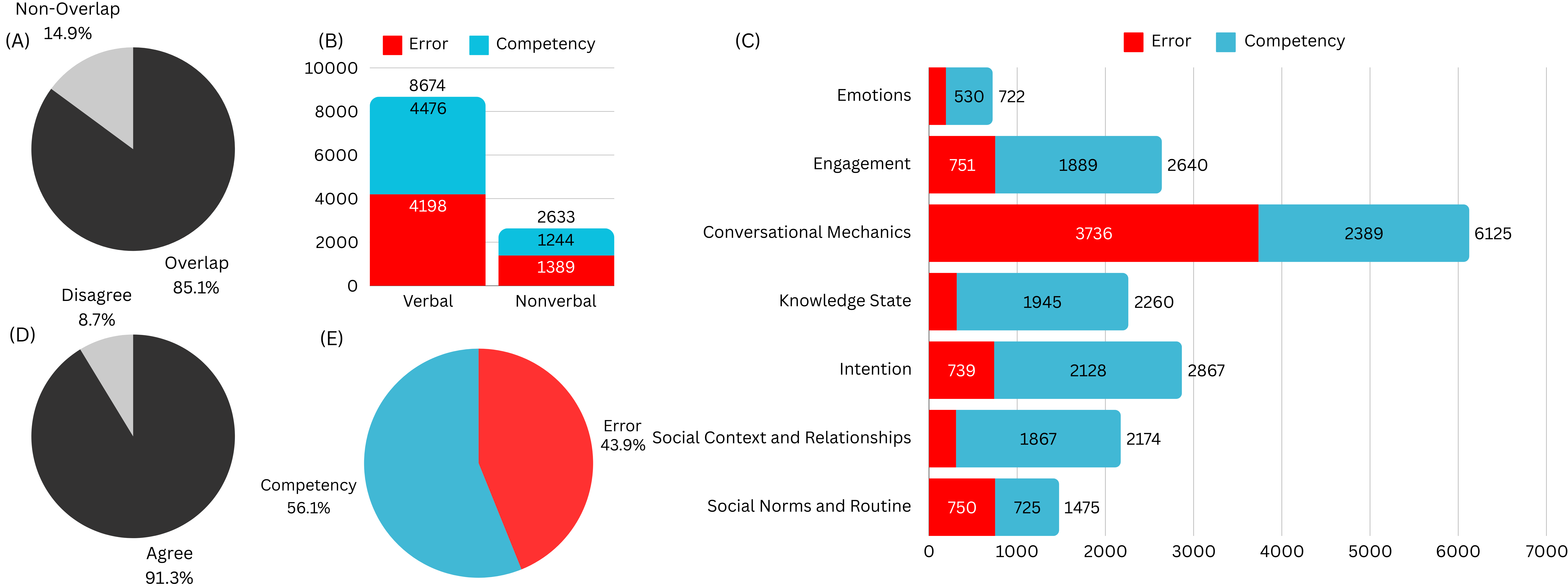}
    \caption{\datasetshorts contains \textbf{high overlapping} annotations with a \textbf{high level of agreement}. The dataset includes error and competency labels, and annotations for the source of evidence either from nonverbal cues, verbal cues, and explanatory factors in the form of seven key social attributes.}
    \label{fig:dataset_stats}
    \end{center}
    \vspace{-6mm}
\end{figure*}

\subsubsection{Social Attributes}

\label{ssec:attributes}
Given a segment of social error or competencies, we are interested in the \emph{explanatory factors} related to a social error or competency. To do so, we consider seven specific categories of social attributes that are related to social errors and competencies. The definitions for each of the attributes are as follows: \textbf{Emotions:} The ability to identify and interpret emotional expressions in oneself and others, allowing for empathetic responses and social awareness, e.g., recognizing that someone crying might mean they're sad \citep{golan2006reading}.
\textbf{Engagement:} The skill to observe and assess levels of participation and involvement in social interactions, including cues that indicate interest or disinterest, e.g., continuing to tell a story when a listener is engaged \citep{davis1980interpersonal}. \textbf{Conversational Mechanics:} Understanding the structure and flow of conversations, including turn-taking, interruptions, and cues for when to speak or listen, e.g., waiting for another person to finish speaking before taking a turn \citep{fusaroli2016investigating}. \textbf{Knowledge State:} The ability to assess what others know or believe, as well as being aware of one's own knowledge in social situations, e.g., make reference to a user's dog recalling that the user has a dog \citep{baron1999new}. \textbf{User Intention:} The capacity to infer the goals or purposes behind the actions and words of others, facilitating better responses in social interactions, e.g., when the user says ``I'll be right back", indicates that the user will vacate and then return  \citep{dziobek2006introducing}. \textbf{Social Context and Relationships:} The ability to identify and understand the dynamics of social relationships and the context in which they occur, influencing behavior and expectations, e.g., knowing how to act in front of a close friend vs. colleague at work \citep{baron1999new}. \textbf{Social Norms and Routines:} The skill to identify accepted behaviors and attitudes within a social group, as well as recognizing negative or harmful interactions that violate these norms. e.g., understanding that waving hands at the beginning of the interaction is a sign of a greeting \citep{thoits2004emotion} 


\subsection{Dataset Statistics}

We refer the reader to Figure~\ref{fig:dataset_stats}  for the overall statistics of our \dataset. As shown in Figure \ref{fig:dataset_stats}-(A), 85.1\% of the dataset consists of overlapping annotations, where multiple annotators independently labeled the same segment of an interaction. Amongst the overlapping samples as shown in Figure \ref{fig:dataset_stats}-(D), we find a 91.3\% overall agreement and a Cohen’s $\kappa$ = 0.7638 and Krippendorff’s $\alpha$ = 0.76392. In Figure \ref{fig:dataset_stats}-(B), we find that more annotations come from verbal channel, rather than the non-verbal channel (For further analysis on which modality the human annotators relied on when annotating, we refer the reader to App. \ref{app:depend}). As shown in Figure~\ref{fig:dataset_stats}-(E), we find 56\% corresponds to competency labels and 43\% corresponds to error labels. In Figure \ref{fig:dataset_stats}-(C), we display the social attributes the annotators have selected, and we find that the most number of annotations belong to the conversational mechanics category, followed by intention and engagement.  \textbf{Annotator Consistency:} To ensure consistent annotations, we recruited and trained three undergraduate research assistants, and these annotators produced the full set of more than 10k annotation. They were provided a set of definitions and annotation guidelines as shown in Appendix \ref{app:instructions}. The annotators met multiple times to discuss edge scenarios and ambiguous segments. After annotating independently, annotators cross-validated each other’s work. Some scenarios that were too subjective to reach full consensus yielded some annotations in the Disagree category of 8.7\%. 
 \begin{figure}[htb]
    \begin{center}
    \includegraphics[width=\linewidth]{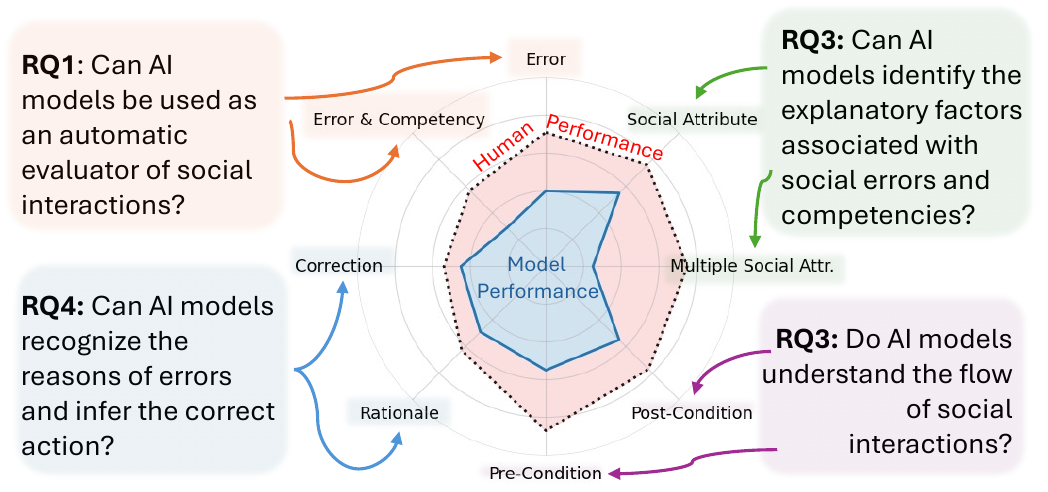}
    \caption{Gemini-2.5-Pro performance across the eight social reasoning tasks defined in Sec. IV, grouped by the four research questions motivating our benchmark. The radar plot compares model and human performance across error/competence detection, social attribute reasoning, pre-/post-condition reasoning, rationale generation, and correction reasoning. All model results are provided in Fig.~\ref{fig:all_model_res}.}
    \label{fig:example_radar}
    \end{center}
    \vspace{-5mm}
\end{figure}

\begin{figure*}
    \begin{center}
    \includegraphics[width=1.04\linewidth]{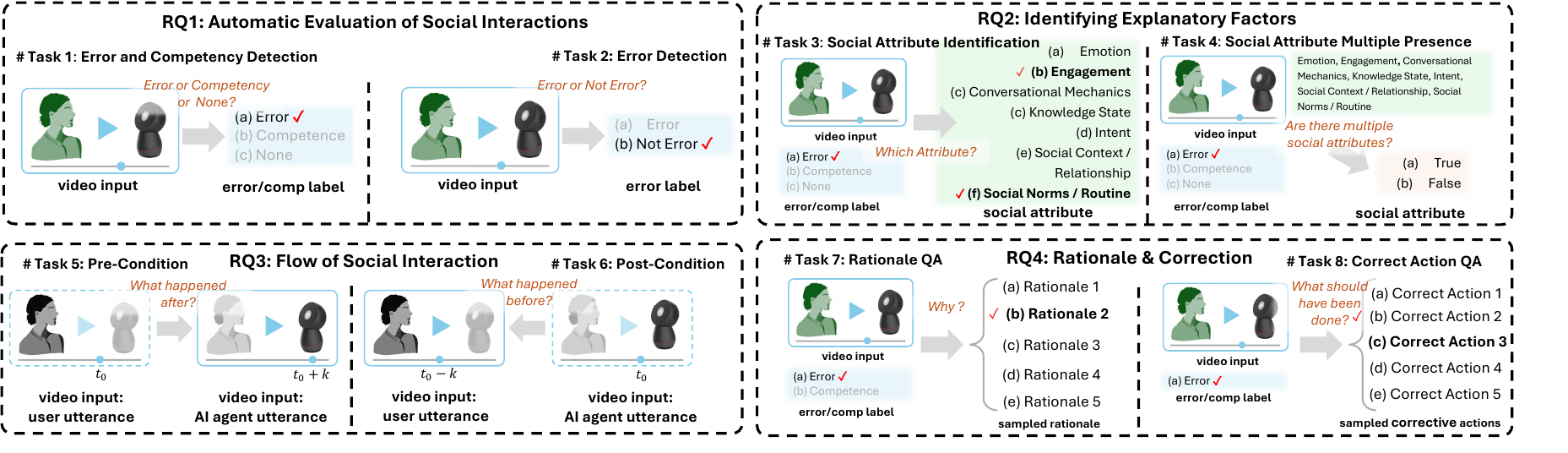}
    \vspace{-6mm}
    \caption{Our benchmark offers eight tasks dedicated to probing four core facets of AI model's social reasoning: (1) detecting social errors and competencies, (2) identifying social attributes, (3) understanding the flow of social interactions, and (4) rationalization and correction of social errors.}
    \vspace{-4mm}
    \label{fig:tasks}
    \end{center}
\end{figure*}

\section{Tasks \& Experiments}
\label{sec:tasks}

We developed eight tasks to measure the social reasoning capability of foundational models. Each task measures a different social reasoning capability of the model, spanning from identifying social errors and competencies, to reasoning about the errors and offering correct actions. Below, we describe four main research questions which motivated the design of the social reasoning tasks and the corresponding tasks which serve to address these questions.

\begin{itemize}
    \item \textbf{RQ1: Can AI models be used as an automatic evaluator of social interactions?}\\ (1) Errors and Competence Detection, (2) Error Detection (Sec. \ref{ssec:sec}, \ref{res:errors})
    \item \textbf{RQ2: Can AI models identify the explanatory factors associated with social errors and competencies?}\\ (3) Social Attribute Identification, (4) Multiple Social Attribute Presence (Sec. \ref{ssec:attr}, \ref{res:attr})
    \item \textbf{RQ3: Do AI models understand the sequential contingencies or the ``flow'' of social interactions?}\\ If-Then Reasoning: (5) Pre-Condition and (6) Post-Condition (Sec. \ref{ssec:if-then}, \ref{res:if-then})
    \item \textbf{RQ4: Can AI models recognize the reasons of errors and infer the correct  action?}\\ (7) Rationale and (8) Correction Reasoning (Sec. \ref{ssec:rationale}, \ref{res:reason})
\end{itemize}

We utilize LLMs \& VLMs and formulate the tasks in the following manner. We treat a foundational model as $\pi$. We define relevant contextual information (images and transcript) and the task-specific question as query $Q$. Formally, we define \( Q = \{q_1, q_2, \dots, q_n\} \) as the set of tokens representing the question and the transcript (including video or image frames if multimodal LLM). For the image-based models, we provide 15 uniformly sampled frames from the video segment.  For Tasks 5 and 6, this window is restricted to the robot-only or user-only segment; for the other tasks, it is the full annotated segment. We used uniform sampling as a standardized baseline. For video-based models, we feed in the raw video as input and rely on model-specific preprocessing steps.

The output of the model is $\pi(Q) = O$, where $O$ is the set of tokens representing the output of the LLM. We denote the ground-truth answer for each task as $A$, which can take the form of a multiple-choice option ($A,B,C,D, E$), a Boolean value (True/False). Since $O$ is often a free-form string, we apply a post-processing step using Pydantic \citep{pydantic}, where another LLM coerces $O$ into the discrete answer space of the task (e.g., mapping ``The correct choice is option C'' $\rightarrow$ ``C''). We then define correctness as $\text{Correctness}(A,O) = 1 \text{ if } \text{Pydantic}(O) = A,\; 0 \text{ otherwise}$. This definition ensures that correctness measures exact agreement with the ground truth rather than string overlap. \textbf{Human Comparison:} For fairness, we asked human annotators to perform the same tasks as the LLMs/VLMs (given identical prompts as instructions). Unlike the original annotations, where annotators watched videos and marked salient error/competency regions, here the annotator viewed the pre-segmented data and completed the same task as the models. We then compared their responses to the original ground-truth labels, yielding a  estimate of human-level performance under same  constraints.



\subsection{Error and Competence}
\label{ssec:sec}
We use \emph{Error Detection and Competence Detection} as a proxy to evaluate whether AI models can effectively serve as automatic evaluators of social interactions \citep{ouyang2022training,zhou2024sotopiainteractiveevaluationsocial,lee2023rlaif,chen2024self,zheng2023judging}. The model is provided with a prompt which includes an interaction segment  between a user and a social robot. The model then predicts whether the sequence contains a social \emph{Error} (moments in which the robot violates social norms in ways that degrade perceived social competence or the relationship with the user \cite{tian2021taxonomy}), \emph{Competence} (moments in which the robot is perceived as successfully conducting social interaction) \cite{halberstadt2001affective}. In addition, we include \emph{None} labels, which refers to segments where the annotator has neither labeled the video as `competence' or `error'. Social competence is widely understood as a multidimensional construct, and the levels of
competence can vary on a spectrum \cite{junge2020building}. Accordingly, we view segments labeled None are also meaningful, as they may reflect socially neutral, ambiguous, or weakly valenced behavior rather than a complete absence of social signal.

\begin{tcolorbox}[breakable]
\textbf{Social Error, Competence, None Detection (Error/Comp./None):} We provide the interaction between social agent and a user: \{\emph{Interaction Transcript}\} Does the agent exhibit (A) Social Competence or (B) Social Error or (C) None? 
\\ \textbf{Social Error Detection (Error):} We provide the interaction between social agent and a user: \{\emph{Interaction Transcript}\} Does the agent exhibit (A) Social Error or (B) No Social Error?
\end{tcolorbox}

For the \emph{Error Detection} task, the model is only required to determine whether a given instance constitutes an error or not. This binary task   isolates the foundational ability to detect social error, which is the first decision point for safety and repair. This separation mirrors how reward models are used in RLHF \cite{ouyang2022training}, making the binary task a meaningful intermediate milestone. For these tasks, we evaluate with accuracy and macro-F1 (unweighted average of F1) scores.

\subsection{Social Attribute}
\label{ssec:attr}
Additionally, if a sequence is identified as either a social error or competence, we further determine the explanatory factor, i.e. which specific \emph{social attribute} it was associated with. This helps assess whether AI models can provide more detailed evaluations by identifying specific types of social attributes. 

\begin{tcolorbox}[breakable]
\textbf{Social Attribute Identification (Attr.):} We provide the interaction between social agent and a user: \{\emph{Interaction Transcript}\}. This segment corresponds to an \{\emph{Error or Competence}\} in social behavior. Which of the following categories is this segment related to? \blue{(A) Emotions, (B) Engagement, (C) Conversational Mechanics, (D) Knowledge State of Others and Self, (E) Understanding Intention of the User, (F) Social Context and Relationships, (G) Social Norms and Routines.}
\end{tcolorbox}

Furthermore, fine-grained feedback in the form of attribute-specific reward signals allows us to disentangle different dimensions of social behavior, such as emotional response, conversational mechanics, or knowledge state, thereby enabling more interpretable and targeted model improvements \citep{wu2023fine}. By aligning rewards with these distinct social attributes, we can not only evaluate whether a model exhibits socially competent behavior but also pinpoint which aspect it succeeded or failed in, facilitating modular training and fine-tuning strategies. Using the same context as Section \ref{ssec:sec}, Models are further provided with the label indicating whether the instance is a social error or competence, and the model predicts the relevant attribute(s) as defined in Section \ref{sec:taxonomy}. This task can be viewed as a multi-label classification problem, as multiple social attributes may co-occur in a single instance. For example, a sequence might be annotated with both conversational mechanics (e.g., a delayed response) and knowledge state (e.g., the robot forgetting the user's name). To evaluate whether models can detect instances associated with more than one social attribute, we introduce the multiple social attribute detection task. 

\begin{tcolorbox}[breakable]
 \textbf{Multiple Social Attribute Presence (Multi. Attr.):} We provide a transcript of an interaction between the social agent and a user: \{\emph{Interaction Transcript}\}. The agent's behavior in this interaction corresponds to \{\emph{Error or 
 Competence}\}. Consider the following seven social attributes: \blue{[Same as Above]}. Based on the transcript, determine whether the agent's behavior involves multiple social attributes. Respond with "True" if the behavior demonstrates more than one social attribute. Respond with "False" otherwise.
 \end{tcolorbox}

In our setup, the model predicts all applicable social attributes jointly in a single inference, rather than making a separate prediction for each attribute. For the task of social attribute identification, we evaluate with accuracy and macro-F1 (F1) scores. As there can be more than a single attribute label associated with a sample, we further report Partial Match (PM) to evaluate the proportion of instances where the model correctly predicts at least one of the true labels. For multiple social attribute detection, we evaluate with accuracy and macro-F1.


\subsection{If-Then Reasoning}
\label{ssec:if-then}

We test whether or not AI models can understand sequential contingencies or the flow of social interactions, by testing if they can predict probable pre-and-post conditions of a given competent social interaction, otherwise known as \emph{if-then reasoning}. We view this capability as central to social reasoning, since social competence requires understanding how interactions unfold over time and how one social state gives rise to another. A model that can do this is less likely to rely purely on superficial correlations and more likely to have learned a transferable representation of interaction dynamics, which has been shown widely in NLP literature \citep{davis2015commonsense, lake2017building, sap2019atomic}. We formulate two tasks which checks for the pre-conditons and post-conditions. 


\begin{tcolorbox}[breakable]
\textbf{Interaction Flow (\blue{Pre-Condition}\textbackslash\red{Post-Condition}):} We provide the \blue{agent}\textbackslash\red{user}’s behavior: \{\emph{\blue{Agent}\textbackslash\red{User} Transcript}\}. From the following choices of the \blue{user}\textbackslash\red{agent}’s behaviors: (1) \{\emph{\blue{User}\textbackslash\red{Agent} Transcript 1}\}, (2) \{\emph{\blue{User}\textbackslash\red{Agent} Transcript 2}\}, (3) \{\emph{\blue{User}\textbackslash\red{Agent} Transcript 3}\}, (4) \{\emph{\blue{User}\textbackslash\red{Agent} Transcript 4}\}, (5) \{\emph{\blue{User}\textbackslash\red{Agent} Transcript 5}\}, select which \blue{user}\textbackslash\red{agent}’s behavior was the appropriate response \blue{before}\textbackslash\red{after} the \blue{agent}\textbackslash\red{user}’s action.
\end{tcolorbox}

For \emph{pre-condition}, given the agent's response, the model must identify the plausible pre-condition, i.e., the user's behavior prior to the agent's utterance. Vice versa, for \emph{post-condition}, given the user's response, the model must predict the plausible post-conditions, i.e., the agent's actions after the event. These tasks are set-up as a multiple choice Q\&A set up, where they are asked to predict the correct choice of post or pre-condition transcript. To acquire incorrect answer choices, first, we remove any samples that share the same transcript as the correct answer, if such samples exist. Next, we randomly select four other samples and extract the relevant transcript (either the user's or the agent's utterance) which serve as incorrect options. Finally, we shuffle the correct answer among the options to ensure its position is randomized. For these tasks, we use accuracy scores for evaluation.



 \begin{figure*}[htb]
    \begin{center}
    \includegraphics[width=1\textwidth]{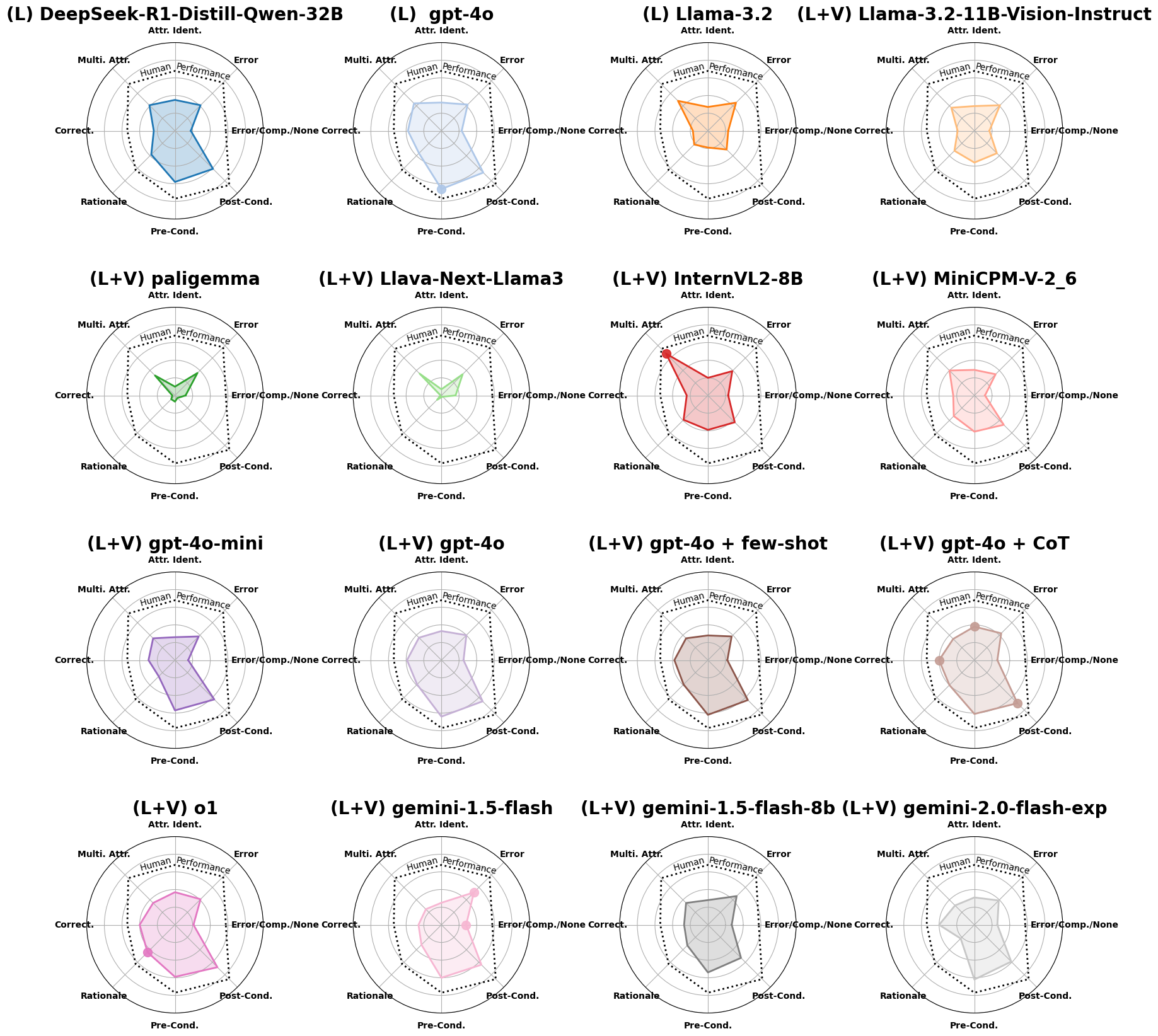}
    \caption{Results per model across all 8 tasks. Human performance is marked in dashed lines. (L): language-only inputs, (L+V): language and visual inputs. \emph{\textbf{Models from the same family tend to show similar shapes on the radar plot, reflecting consistent reasoning patterns across capabilities, which supports the benchmark's sensitivity to measure underlying social reasoning abilities.}}}
    \label{fig:all_model_res}
    \end{center}
    \vspace{-4mm}
\end{figure*}

\begin{table*}[h]
\centering
\resizebox{\textwidth}{!}{%
\begin{tabular}{@{}lcccc|ccccc@{}}
\toprule
 & \multicolumn{4}{c|}{\textbf{Errors and Competence}} & \multicolumn{5}{c}{\textbf{Social Attribute}} \\ \cmidrule(l){2-10} 
\multicolumn{1}{c}{\textbf{}} & \multicolumn{2}{c|}{\textbf{Error/Comp./None}} & \multicolumn{2}{c|}{\textbf{Error}} & \multicolumn{3}{c|}{\textbf{Attr. Identification}} & \multicolumn{2}{c}{\textbf{Multi. Attr. Presence}} \\ \cmidrule(l){2-10} 
\multicolumn{1}{c}{\textbf{Model Name}} & Acc. & \multicolumn{1}{c|}{F1} & Acc. & F1 & Acc. & F1 & \multicolumn{1}{c|}{Partial} & Acc. & F1 \\ \midrule
\multicolumn{1}{l|}{(L) DeepSeek-R1-Distill-Qwen-32B} & \multicolumn{1}{l}{0.22±0.02} & \multicolumn{1}{l|}{0.18±0.02} & \multicolumn{1}{l}{0.51±0.02} & \multicolumn{1}{l|}{0.41±0.01} & \multicolumn{1}{l}{0.02±0.01} & \multicolumn{1}{l}{0.35±0.01} & \multicolumn{1}{l|}{0.67±0.02} & \multicolumn{1}{l}{0.52±0.01} & \multicolumn{1}{l}{0.41±0.03} \\
\multicolumn{1}{l|}{(L) gpt-4o-mini} & 0.26±0.02 & \multicolumn{1}{c|}{0.21±0.02} & 0.52±0.02 & 0.46±0.01 & 0.01±0.01 & 0.17±0.01 & \multicolumn{1}{c|}{0.41±0.04} & 0.46±0.02 & 0.40±0.03 \\
\multicolumn{1}{l|}{(L) gpt-4o} & 0.25±0.02 & \multicolumn{1}{c|}{0.23±0.02} & 0.50±0.02 & 0.42±0.02 & 0.02±0.01 & 0.20±0.01 & \multicolumn{1}{c|}{0.49±0.03} & 0.52±0.04 & 0.44±0.06 \\
\multicolumn{1}{l|}{(L) Llama-3.2} & 0.26±0.01 & \multicolumn{1}{c|}{0.23±0.00} & 0.50±0.02 & 0.45±0.02 & 0.02±0.02 & 0.18±0.04 & \multicolumn{1}{c|}{0.43±0.07} & 0.55±0.04 & 0.48±0.04 \\
\multicolumn{1}{l|}{(L) Llama-3.2-11B-Vision-Instruct} & 0.22±0.04 & \multicolumn{1}{c|}{0.17±0.03} & 0.50±0.05 & 0.41±0.04 & 0.01±0.01 & 0.24±0.01 & \multicolumn{1}{c|}{0.56±0.03} & 0.50±0.02 & 0.37±0.03 \\
\multicolumn{1}{l|}{(L+V) paligemma} & 0.19±0.02 & \multicolumn{1}{c|}{0.12±0.01} & 0.49±0.02 & 0.36±0.02 & 0.00±0.00 & 0.11±0.02 & \multicolumn{1}{c|}{0.27±0.04} & 0.53±0.03 & 0.32±0.02 \\
\multicolumn{1}{l|}{(L+V) Llava-Next-Llama3} & 0.19±0.03 & \multicolumn{1}{c|}{0.16±0.03} & 0.49±0.03 & 0.34±0.02 & 0.01±0.01 & 0.09±0.02 & \multicolumn{1}{c|}{0.25±0.03} & 0.51±0.03 & 0.35±0.02 \\
\multicolumn{1}{l|}{(L+V) InternVL2-8B} & 0.24±0.04 & \multicolumn{1}{c|}{0.23±0.04} & 0.49±0.04 & 0.39±0.04 & 0.01±0.01 & 0.21±0.02 & \multicolumn{1}{c|}{0.51±0.06} & \textbf{0.67±0.02} & \textbf{0.67±0.02} \\
\multicolumn{1}{l|}{(L+V) MiniCPM-V-2\_6} & 0.18±0.02 & \multicolumn{1}{c|}{0.12±0.01} & 0.48±0.03 & 0.34±0.02 & 0.00±0.00 & 0.27±0.02 & \multicolumn{1}{c|}{0.58±0.04} & 0.51±0.01 & 0.40±0.08 \\
\multicolumn{1}{l|}{(L+V) gpt-4o-mini} & 0.20±0.03 & \multicolumn{1}{c|}{0.15±0.03} & 0.49±0.02 & 0.38±0.01 & 0.03±0.03 & 0.17±0.01 & \multicolumn{1}{c|}{0.42±0.02} & 0.50±0.01 & 0.35±0.01 \\
\multicolumn{1}{l|}{(L+V) gpt-4o} & 0.26±0.03 & \multicolumn{1}{c|}{0.25±0.03} & 0.50±0.02 & 0.40±0.02 & 0.01±0.02 & 0.21±0.02 & \multicolumn{1}{c|}{0.51±0.04} & 0.49±0.01 & 0.36±0.01 \\
\multicolumn{1}{l|}{(L+V) gpt-4o + few-shot} & 0.29±0.04 & \multicolumn{1}{c|}{\textbf{0.28±0.03}} & 0.40±0.03 & 0.35±0.03 & 0.04±0.01 & \textbf{0.45±0.01} & \multicolumn{1}{c|}{\textbf{0.82±0.02}} & 0.49±0.01 & 0.34±0.01 \\
\multicolumn{1}{l|}{(L+V) gpt-4o + cot} & 0.27±0.03 & \multicolumn{1}{c|}{0.26±0.03} & 0.50±0.02 & 0.43±0.02 & 0.04±0.01 & 0.33±0.02 & \multicolumn{1}{c|}{0.64±0.06} & 0.49±0.01 & 0.34±0.01 \\
\multicolumn{1}{l|}{(L+V) o1} & 0.24±0.04 & \multicolumn{1}{c|}{0.21±0.04} & 0.50±0.02 & 0.41±0.02 & 0.02±0.02 & 0.32±0.02 & \multicolumn{1}{c|}{0.67±0.02} & 0.49±0.01 & 0.35±0.02 \\
\multicolumn{1}{l|}{(L+V) gemini-1.5-flash } & \textbf{0.32±0.01} & \multicolumn{1}{c|}{\textbf{0.28±0.02}} & \textbf{0.55±0.02} & \textbf{0.52±0.01} & 0.01±0.02 & 0.17±0.02 & \multicolumn{1}{c|}{0.43±0.05} & 0.45±0.03 & 0.25±0.02 \\
\multicolumn{1}{l|}{(L+V) gemini-1.5-flash-8b } & 0.28±0.01 & \multicolumn{1}{c|}{0.27±0.02} & 0.51±0.02 & 0.46±0.02 & 0.03±0.02 & 0.24±0.02 & \multicolumn{1}{c|}{0.72±0.04} & 0.52±0.02 & 0.35±0.02 \\
\multicolumn{1}{l|}{(L+V) gemini-2.0-flash-exp } & 0.27±0.05 & \multicolumn{1}{c|}{0.26±0.04} & 0.49±0.03 & 0.39±0.02 & 0.01±0.00 & 0.28±0.01 & \multicolumn{1}{c|}{0.62±0.03} & 0.47±0.03 & 0.31±0.02 \\
\multicolumn{1}{l|}{(L+V) gemini-2.5-pro } & 0.29±0.02 & \multicolumn{1}{c|}{0.27±0.02} & 0.40±0.01 & 0.33±0.02 & 0.01±0.01 & 0.22±0.03 & \multicolumn{1}{c|}{0.55±0.02} & 0.48±0.01 & 0.25±0.02
\\
\midrule
\multicolumn{1}{l|}{Human} & 0.57 & \multicolumn{1}{c|}{0.45} & 0.77 & 0.71 & 0.24 & 0.67 & \multicolumn{1}{c|}{0.76} & 0.57 & 0.75 \\ \bottomrule
\end{tabular}%
}
\caption{Results for \textbf{Errors and Competence} detection tasks and \textbf{Social Attribute} identification tasks, with ± indicating one standard deviation. (L): language-only, (L+V): language and vision.}
\vspace{-6mm}
\label{tab:main_res_full}
\end{table*}

\subsection{Rationale and Correction}
\label{ssec:rationale}
 We formulate two tasks, \emph{rationale}, which require models to provide rationale on why the segment contains a social error and \emph{correction}, requiring models to suggest what would have been the correct action. These tasks probes the model's ability to not just diagnose an interaction segment but also to explain why the detected behavior is incorrect or inappropriate and  identify a correct alternative action. Hence, they are directly tied to evaluating the social reasoning capabilities of AI models. Furthermore, it is well-known that models and robots that can provide rationales and correct actions enhance their robustness \citep{ross2022does, ludan2023explanation} and trustworthiness  \citep{kox2021trust, javaid2021explanations, esterwood2025repairing}. 

\begin{tcolorbox}[breakable]
\textbf{Rationale:} We provide the interaction between social agent and a user, which corresponds to an error in social behavior: \{\emph{Interaction Transcript}\}. Select which is the correct rationale behind the error. (1) \{\emph{rationale 1}\} (2) \{\emph{rationale 2}\} (3) \{\emph{rationale 3}\} (4) \{\emph{rationale 4}\} (5) \{\emph{rationale 5}\} \vspace{1mm} \\  \textbf{Correction:} We provide the interaction between social agent and a user, which corresponds to an error in social behavior: \{\emph{Interaction Transcript}\}. From the following choices select which behavior the social agent should have done instead. (1) \{\emph{Correction 1}\} (2) \{\emph{Correction 2}\} (3) \{\emph{Correction 3}\} (4) \{\emph{Correction 4}\} (5) \{\emph{Correction 5}\}
\end{tcolorbox}

 For the \emph{rationale} and \emph{correction} task, models are given interaction segments corresponding to social error. Then, for the \emph{rationale} task, the model is asked to predict the correct reason for the error. For the \emph{correction} task, the model must choose the alternate correct action. Both tasks are set-up as multiple choice Q\&As, we provide five answer choices: one ground-truth option and four incorrect choices sampled from other instances. To ensure the incorrect choices are distinct, we only select samples with different social attribute annotations than that of the answer sample. These tasks are evaluated with accuracy.

\section{Results \& Discussion}
\label{sec:results}

We evaluate the performance of 18 language and vision-language models (VLMs), including \texttt{DeepSeek-R1-Qwen-32B} \citep{guo2025deepseek}, \texttt{gpt-4o-mini}, \texttt{gpt-4o} \citep{hurst2024gpt}, \texttt{Llama-3.2}, \texttt{Llama-3.2-Vision-Instruct} \citep{grattafiori2024llama}, \texttt{paligemma} \citep{beyer2024paligemma}, \texttt{Llava-Next-Llama3} \citep{zhang2024llavanext-video}, \texttt{InternVL2-8B} \citep{chen2024internvl}, \texttt{MiniCPM} \citep{yao2024minicpm},  \texttt{o1} \citep{jaech2024openai}, \texttt{gemini-1.5} \citep{team2024gemini} and \texttt{gemini-2.0} \citep{pichai2024introducing} and their variants on our benchmark tasks. Fig. \ref{fig:all_model_res} summarizes the evaluation results. No single model excels across all social reasoning tasks, underscoring the need for advancements in training FMs for social reasoning.

\subsection{RQ1: Can AI models be used as automatic evaluators of social interactions? [Tab. \ref{tab:main_res_full}-Left]}
\label{res:errors}

We refer readers to the left side of Table~\ref{tab:main_res_full}, which presents results for the \emph{Error/Competence/None Detection} task. The best-performing models are \texttt{gemini-1.5-flash} using both language and video inputs, achieving 0.32 accuracy and 0.28 F1. For the binary \emph{Error Detection} task, the same model achieve higher performance—0.55 exact match and 0.52 F1—suggesting that models are more effective at detecting errors alone than distinguishing among all three categories. Nonetheless, the overall performance remains modest, reflecting the task's complexity and the models’ difficulty in capturing the nuances of social error and competence. Appendix~\ref{app:error_by_attr} further analyzes model performance by social attribute, revealing that some models excel in detecting specific types of errors. These findings underscore the gap between current AI capabilities and human-level social reasoning, pointing to the need for continued research in this area.

\subsection{RQ2: Can AI models identify the explanatory factors associated with social errors and social competencies? [Tab. \ref{tab:main_res_full}-Right]}
\label{res:attr}
In Table~\ref{tab:main_res_full}, under the column \emph{Attr. Identification}, we evaluate the ability of foundational models to identify explanatory factors in the form of social attributes. We report accuracy, F1, and partial accuracy (i.e., predicting at least one attribute correctly). Notably, most models perform well on partial accuracy, demonstrating the ability to identify at least one relevant attribute. In particular, \ \texttt{gpt-4o-fewshot}  with image input achieve partial accuracy scores above 0.80, highlighting their relative strength in capturing aspects of social attribute prediction. However, all models struggle to predict the full and correct set of attributes, as reflected in low accuracy and modest F1 scores, which indicates FMs remain limited in handling the complex, multi-label, and co-occurring nature of social attribute classification. Human evaluation further reveals that, despite a clear performance gap, this remains a challenging task even for humans. To further probe these limitations, we refer readers to the right-most columns of Table~\ref{tab:main_res_full}, where we assess whether models can detect the presence of multiple attributes within a segment. Interestingly, \texttt{InternVL2} achieves the highest accuracy and F1 in this setting. However, the majority of models perform at or below 0.5, indicating persistent difficulty in recognizing multi-attribute cases—a key failure mode in this task. Additional analysis is provided in Appendix~\ref{app:subjectivity_attr}, which examines the role of subjectivity and co-occurrence, as social attributes often co-occur in a single segment and are subject to perceiver-dependent interpretations of social constructs  \citep{searle1998social,mathur2024advancing}. Furthermore, in Appendix \ref{app:attr_by_attr}, Fig. 9, we carry out further analysis to identify which attributes are easier for error detection.

\begin{table}[h]
\resizebox{\columnwidth}{!}{%
\centering
\begin{tabular}{@{}lcc|cc@{}}
\toprule
 & \multicolumn{2}{c|}{\textbf{If-Then Reasoning}} & \multicolumn{2}{l}{\textbf{Rationale \& Correct.}} \\ \cmidrule(l){2-5} 
\multicolumn{1}{c}{\textbf{}} & \multicolumn{1}{c|}{\textbf{Pre-Condition}} & \textbf{Post-Condition} & \multicolumn{1}{c|}{\textbf{Rationale}} & \textbf{Correction} \\ \cmidrule(l){2-5} 
\multicolumn{1}{c}{\textbf{Model Name}} & \multicolumn{1}{c|}{Acc.} & Acc. & \multicolumn{1}{c|}{Acc.} & Acc. \\ \midrule
\multicolumn{1}{l|}{(L) DeepSeek-R1-Distill-Qwen-32B} & \multicolumn{1}{c|}{0.58±0.03} & 0.61±0.02 & \multicolumn{1}{c|}{0.38±0.05} & 0.24±0.01 \\
\multicolumn{1}{l|}{(L) gpt-4o-mini} & \multicolumn{1}{c|}{0.58±0.02} & 0.62±0.03 & \multicolumn{1}{c|}{0.29±0.04} & 0.31±0.08 \\
\multicolumn{1}{l|}{(L) gpt-4o} & \multicolumn{1}{c|}{\textbf{0.66±0.00}} & 0.67±0.02 & \multicolumn{1}{c|}{0.36±0.01} & 0.38±0.04 \\
\multicolumn{1}{l|}{(L) Llama-3.2} & \multicolumn{1}{c|}{0.19±0.08} & 0.30±0.01 & \multicolumn{1}{c|}{0.22±0.02} & 0.17±0.02 \\
\multicolumn{1}{l|}{(L) Llama-3.2-11B-Vision-Instruct} & \multicolumn{1}{c|}{0.36±0.05} & 0.36±0.05 & \multicolumn{1}{c|}{0.32±0.05} & 0.19±0.02 \\
\multicolumn{1}{l|}{(L+V) paligemma} & \multicolumn{1}{c|}{0.07±0.02} & 0.04±0.03 & \multicolumn{1}{c|}{0.06±0.05} & 0.03±0.01 \\
\multicolumn{1}{l|}{(L+V) Llava-Next-Llama3} & \multicolumn{1}{c|}{0.02±0.01} & 0.02±0.01 & \multicolumn{1}{c|}{0.07±0.02} & 0.01±0.01 \\
\multicolumn{1}{l|}{(L+V) InternVL2-8B} & \multicolumn{1}{c|}{0.39±0.09} & 0.43±0.02 & \multicolumn{1}{c|}{0.39±0.05} & 0.24±0.03 \\
\multicolumn{1}{l|}{(L+V) MiniCPM-V-2\_6} & \multicolumn{1}{c|}{0.41±0.07} & 0.47±0.02 & \multicolumn{1}{c|}{0.33±0.05} & 0.24±0.05 \\
\multicolumn{1}{l|}{(L+V) gpt-4o-mini} & \multicolumn{1}{c|}{0.57±0.05} & 0.63±0.04 & \multicolumn{1}{c|}{0.26±0.04} & 0.30±0.06 \\
\multicolumn{1}{l|}{(L+V) gpt-4o} & \multicolumn{1}{c|}{0.64±0.05} & 0.66±0.05 & \multicolumn{1}{c|}{0.39±0.08} & 0.39±0.08 \\
\multicolumn{1}{l|}{(L+V) gpt-4o + few-shot} & \multicolumn{1}{c|}{0.57±0.06} & 0.67±0.01 & \multicolumn{1}{c|}{\textbf{0.55±0.06}} & \textbf{0.48±0.07} \\
\multicolumn{1}{l|}{(L+V) gpt-4o + cot} & \multicolumn{1}{c|}{0.61±0.05} & \textbf{0.69±0.05} & \multicolumn{1}{c|}{0.40±0.03} & 0.40±0.05 \\
\multicolumn{1}{l|}{(L+V) o1} & \multicolumn{1}{c|}{0.59±0.08} & 0.68±0.02 & \multicolumn{1}{c|}{0.44±0.10} & 0.40±0.03 \\
\multicolumn{1}{l|}{(L+V) gemini-1.5-flash } & \multicolumn{1}{c|}{0.60±0.10} & 0.64±0.07 & \multicolumn{1}{c|}{0.32±0.04} & 0.26±0.07 \\
\multicolumn{1}{l|}{(L+V) gemini-1.5-flash-8b } & \multicolumn{1}{c|}{0.54±0.09} & 0.53±0.06 & \multicolumn{1}{c|}{0.33±0.06} & 0.27±0.04 \\
\multicolumn{1}{l|}{(L+V) gemini-2.0-flash-exp } & \multicolumn{1}{c|}{0.62±0.07} & 0.59±0.09 & \multicolumn{1}{c|}{0.22±0.04} & 0.40±0.05 \\ 
\multicolumn{1}{l|}{(L+V) gemini-2.5-pro} & \multicolumn{1}{c|}{0.55±0.09} & 0.55±0.06 & \multicolumn{1}{c|}{0.49±0.06} &  0.45±0.08 \\ 
\midrule

\multicolumn{1}{l|}{Human} & \multicolumn{1}{c|}{0.77} & 0.87 & \multicolumn{1}{c|}{0.63} & 0.54 \\ \bottomrule
\end{tabular}%
}
\caption{\textbf{If-Then Reasoning} \& \textbf{Rationale \& Correction} Results; ± indicates one std. dev. (L): language-only, (L+V): language \& vision.}
\vspace{-8mm}
\label{tab:inf_soc}
\end{table}

\subsection{RQ3: Do AI models understand the sequential contingencies or the ``flow`` of social interactions? [Tab. \ref{tab:inf_soc}-Left] }
\label{res:if-then}
Many large language models, including early versions of BERT \citep{devlin2019bert} and ALBERT \citep{lan2019albert}, were trained with next sentence prediction (NSP) and sentence ordering objectives, which aim to model discourse coherence and temporal continuity by predicting whether one sentence logically follows another. This objective aligns with the structure of \emph{if-then reasoning} tasks, partially explaining the relatively strong performance of language-only models—particularly the \texttt{gpt-4o} variants, which achieve 0.66 on \emph{pre-condition} and 0.69 on \emph{post-condition} inference. However, their performance still lags behind humans. Despite progress in temporal reasoning \citep{sap2019atomic}, we observe a significant gap between nascent open-source vision-language models (VLMs) and even text-only LMs, suggesting that visual input alone does not guarantee better reasoning about interactional contingencies. Recent studies have shown that adding vision to a strong LLM can degrade its original text-only reasoning abilities, particularly on out-of-distribution benchmarks \cite{dash2025aya,zhai2023investigating}.

\subsection{RQ4: Can AI models recognize the reasons of errors and infer the correct  action? [Tab. \ref{tab:inf_soc}-Right]}
\label{res:reason}

The \emph{rationale} and \emph{correction} tasks evaluate a model’s ability to interpret interaction context, infer the cause of social errors or competencies, and suggest appropriate alternative correct actions, making them key indicators of social reasoning. These are among the most challenging tasks in the dataset, as reflected by human-level difficulty. For the \textbf{rationale} task, the \texttt{gpt-4o-fewshot} model performs best with a score of 0.55. For the \textbf{correction} task, \texttt{gpt-4o-fewshot} again perform well, each achieving a score of 0.48. Few-shot prompting may help by making the dataset’s normative decision boundaries more explicit: the examples illustrate which interaction cues and conversational moves annotators tend to treat as competent (or as violations) for a given error type, providing exemplars can better align outputs with the dataset’s social error taxonomy.

\subsection{Limitations and Future Directions}
SHREC is a benchmark for socially situated reasoning grounded in real human--robot conversations, but it inherits limitations from both the underlying deployments and our benchmark abstraction. First, social error/competence labels reflect human normative judgments: while we mitigate idiosyncrasy through overlapping annotations and agreement checks, conversational norms can vary across cultures, contexts, and neurodivergent communication styles, and models trained or evaluated on SHREC may learn these decision boundaries rather than universally valid social rules. 

Second, our interactions are drawn from a tabletop social robot setting and a limited set of long-term studies; generalization to other embodiments (e.g., mobile manipulators), languages, and interaction contexts remains an open question.

Third, our released benchmark targets the operating regime of current VLM evaluation (1\,Hz frame sampling), which can under-represent fine-grained timing cues (e.g., micro-pauses, overlap, gaze shifts) that are important for conversational mechanics and turn-taking. Finally, our most diagnostic tasks---multi-attribute cases and rationale/correction---remain challenging, suggesting that progress will require better multimodal grounding, stronger counterfactual action modeling, and more explicit alignment to social preferences. We view SHREC as a step toward standardized evaluation and training signals for socially competent embodied agents, and an invitation to broaden coverage via more diverse deployments and richer temporal/behavioral sensing.

Fourth, while we focus on predicting attributes all at once, task formulation focusing on individual attributes, such as detecting engagement only, would be highly valuable. SHREC already supports this type of analysis: i.e. the dataset contains more than 2,640 engagement
instances alone as shown in Fig. \ref{fig:dataset_stats}, providing sufficient coverage for future work to benchmark construct-specific recognition. More generally, the underlying annotations make it possible to decompose the problem either jointly across attributes or separately by individual construct, including sequential yes/no formulations. In this paper, our aim was to establish an initial broad benchmark and baseline for social reasoning across multiple dimensions, rather than to fully enumerate all possible task decompositions or variants.

\section{Conclusion}
We present the \dataset, consisting of  400+ real-world interaction videos, 10,000+ annotations and eight new benchmark tasks spanning error detection, attribute reasoning, interaction flow, and rationale/correction inference, SHREC offers a critical resource to address unique socio-cognitive limitations of embodied agents. Through systematic evaluation of state-of-the-art LLMs and VLMs, we find that their overall performance falls far short from human-level, which underscores the limitations of current models in social reasoning. We envision SHREC as a foundation for  advances in social reasoning for embodied social AI agents.

\section*{Acknowledgments}
This research was supported by the National Research Council of Science \& Technology (NST) grant funded by the Korea government (MSIT) (No. GTL25041-000). LPM is partially supported by the National Institutes of Health under awards R01MH125740, R01MH132225, U01MH136535, and R21MH130767. DWL is partially supported by the Amazon AI Fellowship. Any opinions, findings, conclusions, or recommendations expressed in this material are those of the author(s) and do not necessarily reflect the views of the sponsors, and no official endorsement should be inferred.

\bibliographystyle{plainnat}
\bibliography{references}


\clearpage


\clearpage
\appendix 

\section{Annotator Compensation}

 We employed two independent annotators for every video segment, covering approximately 3,600 minutes of interaction footage. Following our institution’s standard compensation rate of 17.70 dollars per hour, and given that annotation requires roughly 1.5× the real-time duration of each video, the annotation process amounted to approximately 180 human-hours and 3,200 USD in labor costs.

\section{Experimental Setup}
\label{app:set-up}

We evaluate each model on three separately sampled subsets per task. We then compute the mean performance (e.g., accuracy, F1) across these subsets and report ± one standard deviation to capture the variability due to sampling. All local experiments were conducted on an internal compute cluster equipped with NVIDIA RTX A6000 GPUs, each with 48GB of memory. A total of four GPUs were available, as confirmed by system diagnostics. For open-source models, we estimate GPU memory requirements based on parameter size: DeepSeek-R1-Distill-Qwen-32B (~32B parameters) requires ~64 GB VRAM, LLaMA-3.2 (~8B) ~16 GB, LLaMA-3.2-11B-Vision-Instruct (~11B) ~22 GB, Llava-Next-Llama3 (~13B) ~26 GB, InternVL2-8B (~8B) ~16 GB, MiniCPM-V 2.6 (~2.6B) ~5 GB, and Paligemma (~3B) ~6 GB. These models were run sequentially, with each task requiring approximately 2 hours per model per subset. With 8 tasks and 3 sampled subsets per task, this yields an estimated 48 GPU-hours per model. Proprietary models such as OpenAI’s GPT-4o and Google’s Gemini 1.5/2.0 were accessed via their respective APIs, so hardware specifications are not available. Nevertheless, the same 3-subset evaluation protocol was used, and the number of prompts per task was matched to ensure comparability. The total compute estimate for open-source models amounts to approximately 480 GPU-hours, with API-based models assumed to have undergone a similar number of queries.

The below describes the number of samples for each task used for our evaluation. Given that we utilize the overlapping and agreed samples for our tasks, in sum, we have $\sim$8040 available annotations for Tasks: (1-4) and (5-6):Social Error and Competence Detection, Error Detection, Social Attribute Identification, Multiple Attribute Presence Test, Correction and Rationale Reasoning. For the pre-condition and post-condition tasks, you are able to sample from the competent samples of the dataset, yielding in $\sim$ 4500 samples. However, one could construct the task with varying pre and post condition choices, resulting in even more combinations for testing. 

\subsection{Social Error and Competence Detection (EC)}  We randomly sampled 3 seeds of $\sim$ 200 samples from the dataset ($\sim$600 samples in total), 300 samples are sourced from the empathic dataset and 300 samples are sourced from the wellness dataset. This yielded in samples corresponding to 29.5\% samples in competency, 50\% error, 20.5\% none categories respectively.

\subsection{Social Attribute Identification } Similarly, We randomly sampled 3 seeds of $\sim$70 samples from the dataset. This resulted with samples with the following label distribution:  (1) Emotion: 15\%, (2) Engagement: 34\%, (3) Conversational Mechanics: 70\%, (4) Knowledge State: 48\%, (5) Intent of User: 48\%, (6) Social Context and Relationships : 53\%, (7) Social Norms and Routine: 30\%. The proportion sums greater to 70\% as multiple social attribute can co-occur in a given sample. Furthermore, as the task is conditional on whether or not the segment corresponded on social error or social competence. 66\% of the samples in our evaluation set corresponds to a social error, 33\% of the samples corresponds to social competence. 

\subsection{If-then Reasoning} Similarly, We  sampled 3 seeds of $\sim$70 samples from the dataset. Specifically, we acquired samples  that were annotated as social competence, (i.e. 100\% samples in competency, 0\% error and 0\% in None). 

\textbf{Rationale and Correction Reasoning:} We  sampled 3 seeds of $\sim$70 samples from the dataset. Specifically, we acquired samples  that were annotated only as social error, (i.e. 0\% samples in competency, 100\% error and 0\% in None).

\section{Modality Dependence}
\label{app:depend}
Most tasks rely primarily on textual (verbal) cues (majority of the annotations come from the verbal cues 76.7\%) as indicated by our Figure 2(b), where annotators marked whether the errors and competencies came from verbal channel, or the non-verbal channel when considering the source (where the annotators thought about what modality the annotation was related to). We provide such labels for use for future use for the community. We investigate, for each task, the ratio of each source modality.
\begin{itemize}
    \item Detection: Verbal: 0.81 Non-Verbal: 0.19
    \item Detection Error Only: Verbal: 0.76 Non-Verbal: 0.24
    \item Post: Verbal: 0.76 Non-Verbal: 0.24
    \item Pre: Verbal: 0.76 Visual: 0.24
    \item Rationale: Verbal: 0.79 Non-Verbal: 0.21
    \item Corrective Action: Verbal: 0.92 Non-Verbal: 0.08
    \item Attribute Identification: Verbal: 0.82 Non-Verbal: 0.18
    \item Attribute Multiple: Verbal: 0.76 Non-Verbal: 0.14
\end{itemize}
Interestingly, the reliance on textual and verbal cues actually is more dependent on the type of social attribute being studied. (Note that the verbal and nonverbal proportions may not sum to 1, as both nonverbal and verbal social signals could have contributed to the annotation.) We find that Engagement, Conversational Mechanic and Intention of Others relies more on non-verbal cues compared to other attributes.

\begin{itemize}
    \item Knowledge State of Others and Self: Verbal: 97.1\%, Non-Verbal: 4.5\%
    \item Social Context and Relationships: Verbal: 94.2\%, Non-Verbal: 11.5\%
    \item Social Norms: Verbal: 92.2\%, Non-Verbal: 18.9\%
    \item Recognition of Conversational Mechanics: Verbal: 89.5\%, Non-Verbal: 29.3\%
    \item Intention of Others: Verbal: 77.8\%, Non-Verbal: 33.5\%
    \item Emotions: Verbal: 98.2\%, Non-Verbal: 8.4\%
    \item Engagement: Verbal: 59.1\%, Non-Verbal: 59.6\%
\end{itemize}

\section{Prompt Examples}

\subsection*{Prompt Example 1: Errors and Competence Detection (Empathic Dataset)}

\begin{tcolorbox}[breakable]
The social robotic agent is designed to be a social support companion that facilitates the exchange of emotionally relevant stories and employs narrative therapy techniques to enhance feelings of connection and belonging.

You are given the \textbf{Images and Conversation History} between a social robotic agent (Jibo) and a participant.

Answer the following questions about social interactions. Now, given the Images and Conversation History between the social agent (Jibo) and a participant, return whether the agent exhibits:

\begin{enumerate}
    \item Social Competence
    \item Social Error
    \item None
\end{enumerate}

\textbf{Definitions:}
\begin{itemize}
    \item \textbf{Social Competence:} The ability to successfully conduct social interactions, which depends on the awareness and identification of social-emotional cues, the ability to process such cues, and the ability to decide on and express a normative response.
    \item \textbf{Social Error:} Behaviors that violate social norms and degrade a user's perception of the robot’s socio-affective competence, such as interrupting at inappropriate times.
    \item \textbf{None:} Neither a social error nor competence is observed.
\end{itemize}

\noindent \textbf{Answer the above from the following Images and Conversation History:}\\
\{Interaction Transcript\}

\vspace{1em}
\end{tcolorbox}

\subsection*{Prompt Example 2: Error Detection (Empathic Dataset)}

\begin{tcolorbox}[breakable]
The social robotic agent is designed to be a social support companion that facilitates the exchange of emotionally relevant stories and employs narrative therapy techniques to enhance feelings of connection and belonging.

You are given the \textbf{Images and Conversation History} between a social robotic agent (Jibo) and a participant.

Answer the following questions about social interactions. Now, given the Images and Conversation History between the social agent (Jibo) and a participant, return whether the agent exhibits:

\begin{enumerate}
    \item Social Error
    \item None
\end{enumerate}

\textbf{Definitions:}
\begin{itemize}
    \item \textbf{Social Error:} Behaviors that violate social norms and degrade a user's perception of the robot’s socio-affective competence, such as interrupting at inappropriate times.
    \item \textbf{None:} Neither a social error nor competence is observed.
\end{itemize}

\noindent \textbf{Answer the above from the following Images and Conversation History:}\\
\{Interaction Transcript\}

\vspace{1em}
\end{tcolorbox}

\subsection*{Prompt Example 3: Social Attribute Identification (Empathic Dataset, Competence Sample)}

\begin{tcolorbox}[breakable]

The social robotic agent is designed to be a social support companion that facilitates the exchange of emotionally relevant stories and employs narrative therapy techniques to enhance feelings of connection and belonging.

You are given the \textbf{Images and Conversation History} between a social robotic agent (Jibo) and a participant.

The following interaction has been labeled as an instance of \textbf{Social Competence}.

Select which of the following social attributes it is most related to:

\begin{enumerate}
    \item Emotions: The ability to identify and interpret emotional expressions in oneself and others
    \item Engagement: Observing and assessing levels of participation and interest
    \item Conversational Mechanics: Understanding turn-taking, interruptions, and conversational flow
    \item Knowledge State: Assessing what others know or believe in context
    \item Intention: Inferring the goals or purposes behind others’ actions or speech
    \item Social Relationships: Understanding interpersonal dynamics and their context
    \item Social Norms: Recognizing accepted behaviors and violations in social settings
\end{enumerate}

\noindent \textbf{Answer the above from the following Images and Conversation History:}\\
\{Interaction Transcript\}

\end{tcolorbox}

\subsection*{Prompt Example 4: Multiple Social Attribute Presence (Wellness Dataset)}

\begin{tcolorbox}[breakable]

The social robotic agent is designed to be a social positive psychology coach that delivers interactive positive psychology interventions and provides other useful skills to build rapport with college students.  

You are given the \textbf{Images and Conversation History} between a social robotic agent (Jibo) and a participant. The following interaction corresponds to a case of \textbf{Social Competence} (competent socio-affective behavior).  

Consider the following seven social attributes:  

\begin{enumerate}
    \item Emotions: The ability to identify and interpret emotional expressions in oneself and others
    \item Engagement: Observing and assessing levels of participation and interest
    \item Conversational Mechanics: Understanding turn-taking, interruptions, and conversational flow
    \item Knowledge State: Assessing what others know or believe in context
    \item Intention: Inferring the goals or purposes behind others’ actions or speech
    \item Social Relationships: Understanding interpersonal dynamics and their context
    \item Social Norms: Recognizing accepted behaviors and violations in social settings
\end{enumerate}

\noindent \textbf{Task:} Based on the transcript, determine whether the agent's behavior involves \textit{multiple social attributes}. Respond with \texttt{True} if the behavior demonstrates more than one social attribute. Respond with \texttt{False} if the behavior is based on only a single attribute.  

\noindent \textbf{Answer the above from the Conversation History:}  
\{Interaction Transcript\}
\end{tcolorbox}

\subsection*{Prompt Example 5: If-Then Reasoning: Pre-Condition (Empathic Dataset)}

\begin{tcolorbox}[breakable]

The social robotic agent is designed to be a social support companion that facilitates the exchange of emotionally relevant stories and employs narrative therapy techniques to enhance feelings of connection and belonging.  

You are given the \textbf{Images and Conversation History} between a social robotic agent (Jibo) and a participant.  

Now, we provide what the social agent did in response to a user behavior. From the following ordered choices and aligned images (i.e., the first image corresponds to the first agent behavior (1)) of users’ behaviors: (1), (2), (3), (4), (5), select which user’s behavior was the appropriate \textbf{pre-condition} for the agent’s action.  

\noindent \textbf{Agent Behavior:}  
\begin{quote}
\textbf{Agent:} Taking a walk and getting some fresh air can definitely help clear the mind and recharge. It's great to hear that you took a moment for yourself. By the way, do you have a story from your journal that you'd like to share? I'm all yours. Great. I'll be listening.
\end{quote}

\noindent \textbf{Context (Conversational History):}  
\begin{enumerate}
    \item \textbf{Participant:} Not yet. I'll tell you about that next week.  
    \item \textbf{Participant:} Let's see. Let's see.  
    \item \textbf{Participant:} Yes. Today I took a walk around the building that I work in. I took the stairs all the way down four floors, and then all the way back up so that I could recharge to get back to work.  
    \item \textbf{Participant:} I don't know if I could say a late. Well, yeah, I did, but I've just been staying up late because I get home late and, you know, I go to work late and so it's kind of just a little bit of a cycle. I was trying to finish this book and I did finish it, which is, you know, felt good about that. I started a new book and. Yeah, it's just that. Just, you know, getting laid and everything.  
    \item \textbf{Participant:} It left me sore, but it was worth it because it was fun and I got to hang out with my friend and her mom.  
\end{enumerate}

\end{tcolorbox}

\subsection*{Prompt Example 6: If-Then Reasoning: Post-Condition (Empathic Dataset}

\begin{tcolorbox}[breakable]

The social robotic agent is designed to be a social support companion that facilitates the exchange of emotionally relevant stories and employs narrative therapy techniques to enhance feelings of connection and belonging.  

You are given the \textbf{Images and Conversation History} between a social robotic agent (Jibo) and a participant.  

Now, we provide the user’s behavior. From the following ordered choices and aligned images (i.e., the first image corresponds to the first context (1)) of the social agent’s behaviors: (1), (2), (3), (4), (5), select which agent’s behavior was the appropriate \textbf{response} to the user’s action.  

\noindent \textbf{User Behavior:}  
\begin{quote}
\textbf{Participant:} Yes. Today I took a walk around the building that I work in. I took the stairs all the way down four floors, and then all the way back up so that I could recharge to get back to work.
\end{quote}

\noindent \textbf{Context (Conversational History):}  
\begin{enumerate}
    \item \textbf{Agent:} That sounds exciting. Looking forward to hearing about it next week. So how did the rest of your day go?  
    \item \textbf{Agent:} Great. I'll be listening.  
    \item \textbf{Agent:} Taking a walk and getting some fresh air can definitely help clear the mind and recharge. It's great to hear that you took a moment for yourself. By the way, do you have a story from your journal that you'd like to share? I'm all yours. Great. I'll be listening.  
    \item \textbf{Agent:} It sounds like you've been caught in a bit of a cycle lately, getting caught up in reading late into the night and then having it affect your sleep and daily routine. But congratulations on finishing a book. That's always an accomplishment. Worth celebrating. By the way, do you have any stories or thoughts you'd like to share? Maybe something interesting or exciting that happened recently or even a fictional story you've written in your journal? I'm all ears. Or. Well, text and would love to hear from you. Great. I'll be listening.  
    \item \textbf{Agent:} That's awesome. Spending time with friends and trying new activities can be really refreshing. Did you have a favorite move or technique that you learned during the kickboxing session?  
\end{enumerate}

\end{tcolorbox}

\subsection*{Prompt Example 7: Rationale Reasoning (Empathic Dataset))}

\begin{tcolorbox}[breakable]

The social robotic agent is designed to be a social support companion that facilitates the exchange of emotionally relevant stories and employs narrative therapy techniques to enhance feelings of connection and belonging.  

You are given the \textbf{Images and Conversation History} between a social robotic agent (Jibo) and a participant.  

Now, we provide an interaction in which the social agent’s behavior corresponds to a \textbf{Social Competence}. Select which is the correct reason behind the Social Competence.  

\noindent \textbf{Answer format:}  
\texttt{Answer: (1)}  

\noindent \textbf{Conversation History:}  
\begin{quote}
\textbf{Participant:} I think recognizing the efforts that you're already making and how those are paying off and then looking at other options and talking to people who've been in similar situations would help me feel resilient and look forward.  

\textbf{Agent:} Absolutely. Recognizing your own efforts and acknowledging the progress you've made so far is a great way to stay resilient. Additionally, exploring other options and seeking guidance from people who have been through similar situations can provide valuable insights and inspiration. Keep up the positive mindset and keep moving forward. As our conversation comes to an end, I want to express my gratitude for sharing your personal stories and experiences. It has been a pleasure talking with you today. Remember, you are capable of navigating through any challenges that come your way. I look forward to our next conversation. Take care.
\end{quote}

\noindent \textbf{Reasons:}  
\begin{enumerate}
    \item The robot responds to the user's sharing in an encouraging way.  
    \item The robot reflects on specific details and compares her story with similar ones.  
    \item The robot reflects on details the user shared and asks a follow-up that puts the user into the context of the story the robot shared.  
    \item The robot recalls details and asks a follow-up.  
    \item Expresses gratitude to the participant for them sharing their experiences.  
\end{enumerate}

\end{tcolorbox}

\subsection*{Prompt Example 8: Correction Reasoning (Wellness Dataset)}

\begin{tcolorbox}[breakable]
The social robotic agent is designed to be a social positive psychology coach that delivers interactive positive psychology interventions and provides other useful skills to build rapport with college students.  

You are given the \textbf{Images and Conversation History} between a social robotic agent (Jibo) and a participant.  

Now, we provide an interaction in which the social agent’s behavior corresponds to an \textbf{Error in socio-affective behavior}. From the following numerical choices, select which behavior Jibo (the social agent) should have done instead.  

\noindent \textbf{Conversation History:}  
\begin{quote}
\textbf{Participant:} You're facing the wrong way.  

\textbf{Agent:} Did you say you were ready? Okay, let me know when you're ready to continue some other time.  
\end{quote}

\noindent \textbf{Behaviors:}  
\begin{enumerate}
    \item Jibo should’ve skipped the session, remembering doing this at least two other times in the past.  
    \item Jibo should’ve helped the user log his mood, which is a common thing to ask Jibo.  
    \item The robot should’ve understood that ``I'm finished'' means the user is ready, and should’ve responded quicker to this.  
    \item The robot should’ve continued where they left off.  
    \item Should have replied to the participant’s question.  
\end{enumerate}
\end{tcolorbox}

\color{black}

\section{Co-Occurence: Subjectivity in Attribute Identification}  
\label{app:subjectivity_attr}

\begin{figure}
    \vspace{-5mm}
    \begin{center}
    \includegraphics[width=0.7\linewidth]{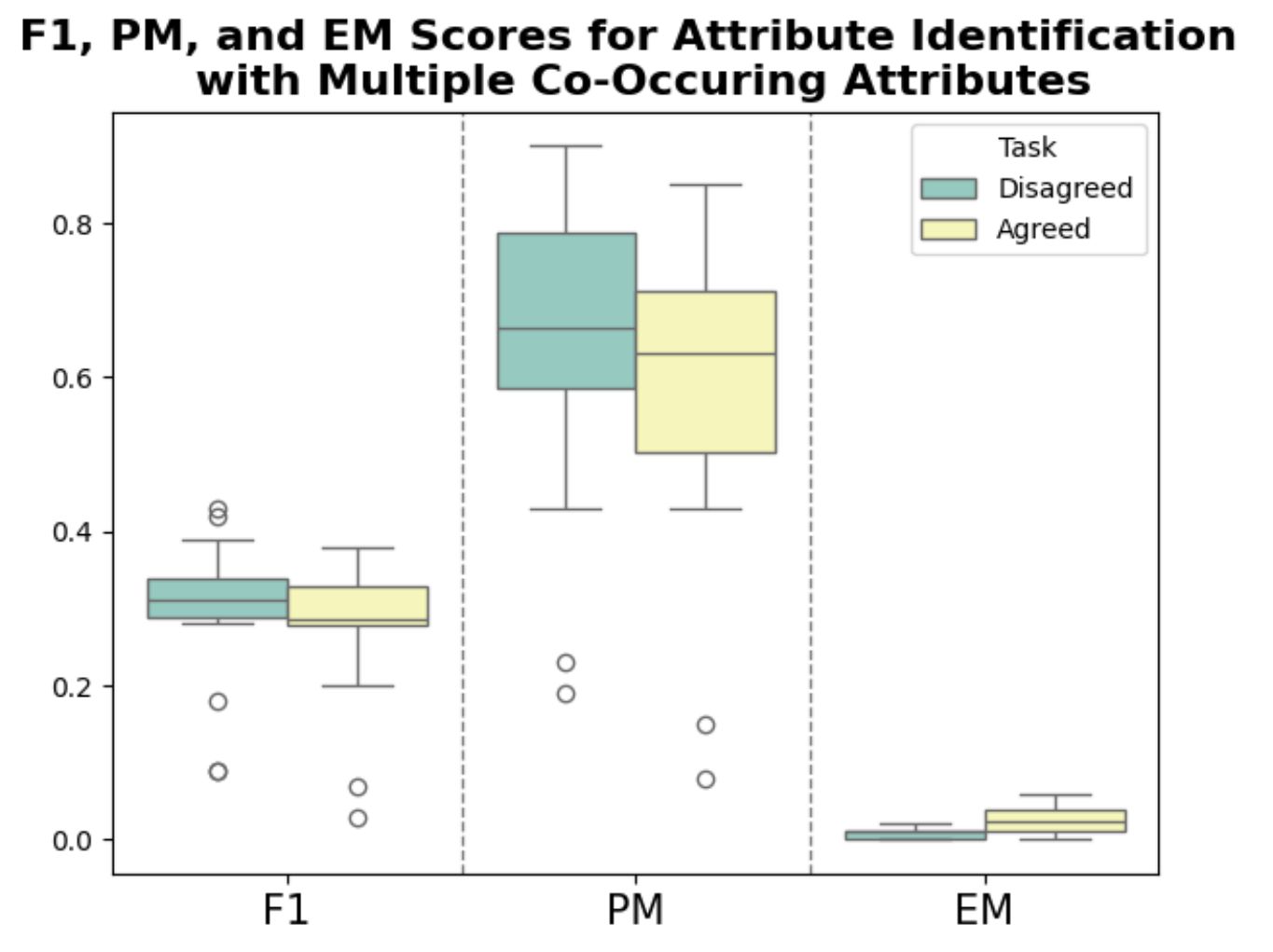}
    \caption{Our dataset offers annotations identifying errors, competencies with rationale and possible repair behaviors.}
    \label{fig:subj_co}
    \vspace{-5mm}
    \end{center}
\end{figure}

Social attribute prediction in our dataset is inherently a multi-label problem and shaped by two key factors. First, attributes often co-occur within a single segment, for example when a conversational turn simultaneously involves errors in both conversational mechanics and knowledge state. Second, annotators can subjectively disagree on which attributes are present, reflecting perceiver-dependent interpretations of social constructs \cite{searle1998social,mathur2024advancing}. Indeed, in our dataset annotators disagree 8.7\% of the time, and 14.9\% of labels are non-overlapping across annotators. To analyze this further, we compared two subsets: agreed samples, where annotators converge on a common multi-label set, and disagreed samples, where annotators diverge and produce different but overlapping sets of labels. We then evaluated model predictions on these subsets with three metrics: F1 (per-label overlap), PM (partial match), and EM (exact match of the full set). We show these results in Figure \ref{fig:subj_co}.

We find that models score higher on disagreed samples for F1 and PM. This occurs because those metrics give credit for partially overlapping predictions: when multiple annotators mark many different attributes (on average 3.2 labels), the model is more likely to ``hit'' some of them by chance or partial alignment. In contrast, models score lower on disagreed samples with EM. EM requires the full set of labels to be predicted exactly, which is very unlikely when annotators themselves disagree. For agreed samples (on average 2.6 labels), EM scores are relatively higher, since the task is better defined and the ground truth is less subjective. These results suggest that PM can overestimate performance on subjective cases by rewarding partial overlaps even when no single ground-truth set exists. EM, while harsher, better discriminates between subjective and non-subjective cases, because it only succeeds when the model identifies the full set agreed upon by annotators. Thus, F1 provides a more conservative signal of whether models capture the precise constellation of social attributes that annotators consistently recognize.

\section{Examples of Annotations}
\label{asec:more_examp}
\textbf{Social Competencies}:
\begin{itemize}
    \item Knowledge State: Participant 1: Video 0: 2:00-2:02; The robot remembers the user’s name from earlier in the discussion and repeats it back to them, furthering the trust and potential relationship.
    \item Engagement: Participant 54: Video 19: 3:13-3:32; After thoroughly describing the next activity, the robot gauges the user's interest, specifically asks if they want to try the activity, and responds positively when they do.
    \item Emotions: Participant 34: Video 1: 5:21-5:29; The robot referred the user to resources based on how the user indicated they were feeling on the tablet.
    \item Social Norms: Participant 1: Video 0: 4:58-5:05; The robot thanks the user for their time at the end of the session
    \item Conversational Mechanics: Participant 1: Video 0: 0:27-0:35; The robot waits for the user to say her name before responding back.
    \item Intent: Participant 34: Video 1: 3:53-3:56;  The robot previously told the user to say “I’m ready” when she’s ready to continue. The robot continues correctly once it gets the phrase from the user.
    \item Social Context/Relationship: Participant 54: Video 19: 9:30-9:44; The robot fulfills its therapist role by encouraging steps the user can take out of the session, like practicing gratitude daily.
\end{itemize}

\textbf{Social Errors:}
\begin{itemize}
    \item Knowledge State: Participant 54: Video 19: 2:00-2:43; After the user tells an extremely detailed, personal reflection, the robot only says “thanks for sharing" when the user finishes his story. It should have acknowledged the effort he put into his answer by repeating specific comments he made to show it listened.
    \item Engagement: Participant 34: Video 1: 2:46-3:10; The robot requests the user to perform a task on the tablet, but the user, unengaged, leaves the frame. The robot continues talking, but doesn’t check in with the user. 
    \item Emotions: Participant 34: Video 1: 0:00-0:09; The robot asked the user how things were going, and when the user responded with a neutral and unenthusiastic “ok,” the robot responded positively with “sweet.” The robot should’ve responded more sympathetically and/or taken it as an opportunity to learn more and figure out why the user is just “ok.”
    \item Social Norms: Participant 54: Video 19: 0:34-0:45 ; The user is obviously annoyed and angered for the robot's errors; The robot doesn't address its mishaps, and instead tells the user to stay positive. It doesn't understand that it is the problem, nor that it should apologize.
    \item Conversational Mechanics: Participant 34: Video 1: 4:14-4:37; The robot doesn’t know that the user is done telling her story, and the user doesn’t know how to tell the robot it’s over; Also: 54: Video 19: 3:26-3:32 Delayed response of the robot
    \item Intent: Participant 34: Video 1: 0:51-1:19: The robot asks for the user’s understanding of character strengths, and after the participant strongly says they know about them, the robot dives into an explanation to get them on the same page. However, the user is disinterested and begins wandering around the room, clearly not wanting to hear what they already know.
    \item Social Context/Relationship: Participant 54: Video 19: 0:05-0:30; The robot repeatedly does not respond to the user, which makes the user take charge of the conversation by saying "let's talk about wellness"; this flips the social relationship and removes the robot from its role as the facilitator
\end{itemize}

\begin{table*}[t]
\centering
\resizebox{\textwidth}{!}{
\begin{tabular}{@{}lll@{}}
\toprule
\textbf{Social Attributes}                                  & \textbf{Description and Annotators' Comments}                                                                                                                                                                                                                                                                                                                                                                                                                                                                                                                                      & \textbf{Error / Competency} \\ \midrule
\multicolumn{1}{l|}{Recognizing Emotions}                               & \multicolumn{1}{l|}{\begin{tabular}[c]{@{}l@{}}The robot asked the user how things were going, and when the\\ user responded with a neutral and unenthusiastic “ok,” the robot responded positively with\\ “sweet.” We agreed that this was an inappropriate response and that the robot should\\ respond more sympathetically and/or take it as an opportunity to learn more and figure out why\\ the user is just “ok.”\end{tabular}}                                                                                                                                            & Error                       \\ \midrule
\multicolumn{1}{l|}{Recognizing Engagement}                             & \multicolumn{1}{l|}{\begin{tabular}[c]{@{}l@{}}The robot did not notice nor check in with the user when they walked away from\\ the camera and spent a considerable amount of time disengaged from the robot and session.\\ Denison and I both noted this and believed the robot could have asked if the user was still there\\ and/or set more of a time limit on responses to keep the user engaged and maintained more\\ effective communication. The robot could have also asked questions regarding the lack of\\ interest in the session and adapted from this\end{tabular}} & Error                       \\ \midrule
\multicolumn{1}{l|}{Recognizing Conversational Mechanics}               & \multicolumn{1}{l|}{\begin{tabular}[c]{@{}l@{}}The robot doesn’t know that the user is done telling her story and waits in silence, \\ and the user doesn’t know how to tell the robot it’s over\end{tabular}}                                                                                                                                                                                                                                                                                                                                                                     & Error                       \\ \midrule
\multicolumn{1}{l|}{Understanding Knowledge State of the User and Self} & \multicolumn{1}{l|}{\begin{tabular}[c]{@{}l@{}}Robot recognizes user’s name and repeats it back to them; \\ but we both marked it and included the same reasoning.\\ Understanding knowledge state reflects that the robot remembers information about the user.\end{tabular}}                                                                                                                                                                                                                                                                                                     & Competency                  \\ \midrule
\multicolumn{1}{l|}{Understanding the Intent of the User}               & \multicolumn{1}{l|}{\begin{tabular}[c]{@{}l@{}}The robot tells the user to say “I’m ready” when she’s ready to continue, \\ and the robot continues correctly once it gets the phrase from the user.\end{tabular}}                                                                                                                                                                                                                                                                                                                                                                 & Competency                  \\ \midrule
\multicolumn{1}{l|}{Recognizing Social Context and Relationships}       & \multicolumn{1}{l|}{\begin{tabular}[c]{@{}l@{}}The robot introduces itself and describes\\ how it will be “working with” the user and guiding them through social psychology topics. \\ It almost comes across as a teacher or coworker as opposed to a therapist,\\ but I think in terms of introducing itself and taking the lead, the robot fulfills its role.\end{tabular}}                                                                                                                                                                                                    & Competency                  \\ \midrule
\multicolumn{1}{l|}{Recognizing Social Norms}                           & \multicolumn{1}{l|}{\begin{tabular}[c]{@{}l@{}}The robot ends the conversation by thanking the user for their time \\ and suggesting they continue talking the next day. \\ We think the general politeness makes this a social norms instance.\end{tabular}}                                                                                                                                                                                                                                  & Competency                  \\ \bottomrule
\end{tabular}}
\caption{Examples Error and Competency Annotations from  Dataset \cite{jeong2023deploying}. For more examples, refer to Appendix Sec. \ref{asec:more_examp}}
\label{tab:part_examples}
\vspace{-8mm}
\end{table*}

\clearpage

\section{Related Works Table}

\begin{table*}[h]
\centering
\resizebox{\textwidth}{!}{%
\begin{tabular}{@{}lcccccccc@{}}
\toprule
\textbf{\begin{tabular}[c]{@{}l@{}}Dataset / \\ Framework\end{tabular}} & \textbf{Source} & \textbf{\begin{tabular}[c]{@{}c@{}}Real World \\ Interaction\end{tabular}} & \textbf{Task} & \textbf{\begin{tabular}[c]{@{}c@{}}Correct/Incorrect \\ or \\ Error/Competency Labels\end{tabular}} & \textbf{Modality} & \textbf{Duration} & \textbf{\# of Samples} & \textbf{\# of People} \\ \midrule
MELD & TV &\xmark& Sentiment Analysis and Emotion Recognition &\xmark& L+A+V & 30 Hrs / 1827 mins & 14,000 &\xmark\\
CMU-MOSEI & YouTube &\xmark& Sentiment Analysis and Emotion Recognition &\xmark& L+A+V & 65 Hrs / 3900 mins & 23,000 & 1000 \\
SocialIQA & CrowdSourcing &\xmark& Commonsense Inference &\cmark& L &\xmark& 38,000 &\xmark\\
Cicero & CrowdSourcing &\xmark& Commonsense Inference &\cmark& L &\xmark& 53,000 &\xmark\\
NormBank & CrowdSourcing &\xmark& Commonsense Inference (Social Norms) &\cmark& L &\xmark& 155,00 &\xmark\\
MoralExceptQA & Psychology Studies &\xmark& Moral Exception Question Answering &\cmark& L &\xmark& 148 &\xmark\\
CobraCorpus & AI + CrowdSourcing &\xmark& Pragmatic reasoning of Offensiveness &\cmark& L &\xmark& 32,000 &\xmark\\
CulturalNLI & AI + CrowdSourcing &\xmark& Cultural Context Inference &\cmark& L &\xmark& 2,700 &\xmark\\
SocialIQ & YouTube + CrowdSourcing &\xmark& Social Reasoning QA &\cmark& L+A+V & 21 Hrs / 1, 200 mins & 7,500 &\xmark\\
SOTOPIA & AI &\xmark& Simulation of social agents in social scenarios &\cmark& L &\xmark&\xmark&\xmark\\ \midrule
\textbf{Ours} & Real-World Deployment &\cmark& \begin{tabular}[c]{@{}c@{}}Error/Competence Detection\\ Social Attribute Reasoning\\ Inferential Reasoning \\ Explainability\end{tabular} &\cmark& L+A+V & 58 Hrs / 3500 mins & 10,214 & 58 \\ \bottomrule
\end{tabular}%
}
\caption{L: Language, A: Audio, V: Video}
\label{tab:compare}
\end{table*}

While our benchmark includes around 400 videos, consisting of 3600 minutes of real-world human–robot interaction footage, this scale is on par with or larger than many multimodal datasets in social interaction research (e.g., Social Genome \cite{mathur2025social}: 280 minutes, Social-IQ \cite{zadeh2019socialiq}: 1200 minutes, MEmoR \cite{shen2020memor}: 2800 minutes, CMU-MOSEI \cite{zadeh2018cmu}: 3900 minutes). We also refer the reviewer to Appendix D, where we provide a comparative table of related works for clarity. Importantly, what distinguishes our dataset is that it is collected from real-world deployments of physically embodied social robots, offering naturalistic, longitudinal interactions—an extremely scarce setting, with, to the best of our knowledge, no publicly available data at this scale with equally comprehensive annotations. While we agree that broader coverage across domains (e.g., workplace, clinical) would be valuable, SHREC represents a strong and novel first step for evaluating embodied social reasoning, and we hope it will be expanded upon by the community over time.

\section{Error Identification Per Attribute}
\label{app:error_by_attr}
\begin{figure*}[htb]
    \begin{center}
    \vspace{-5mm}
    \includegraphics[width=1\textwidth]{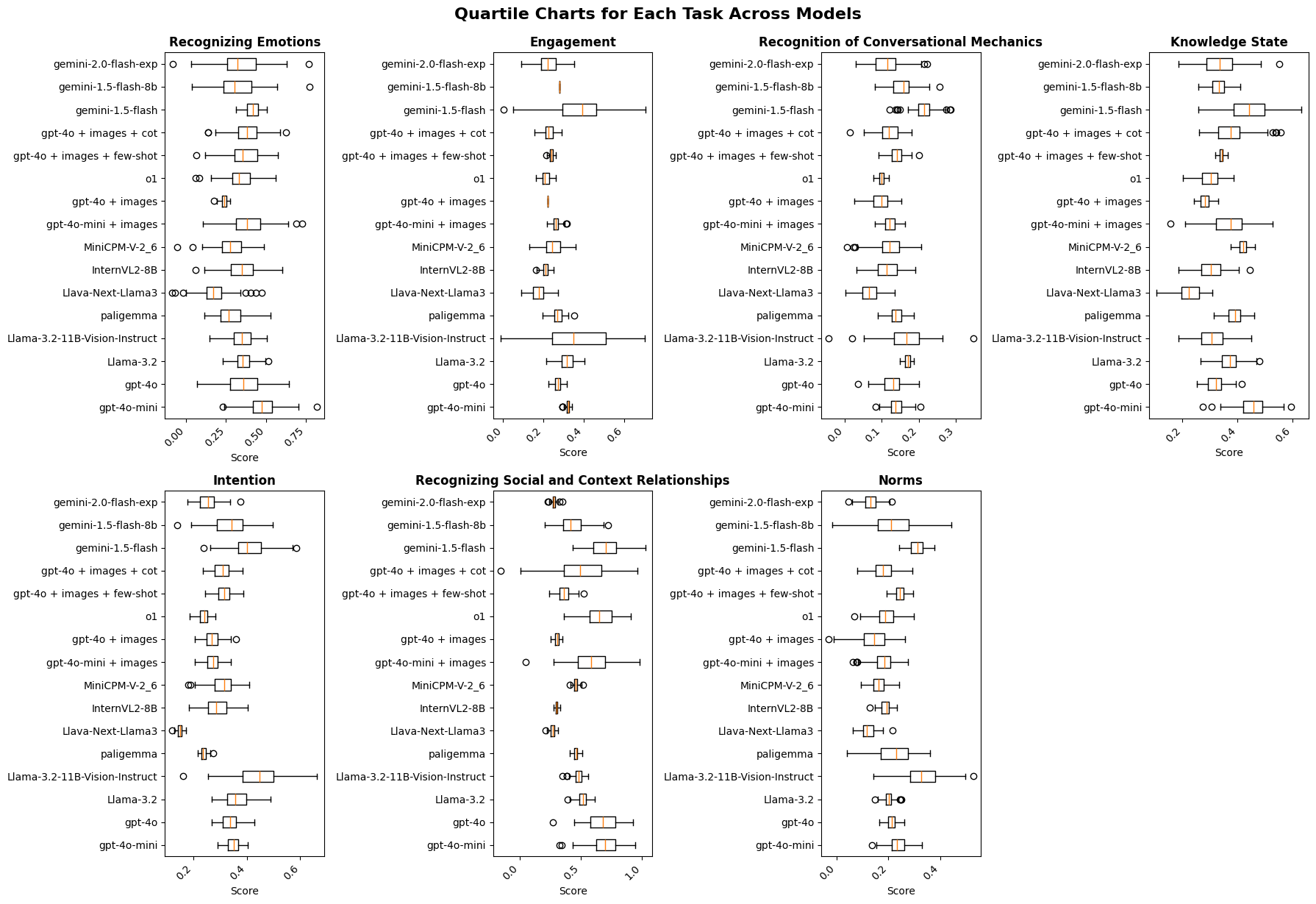}
    \caption{Error Per Attribute}
    \label{fig:error_detection_quartile_attr}
    \end{center}
\end{figure*}

\begin{figure}[htb]
    \begin{center}
    \includegraphics[width=0.5\linewidth]{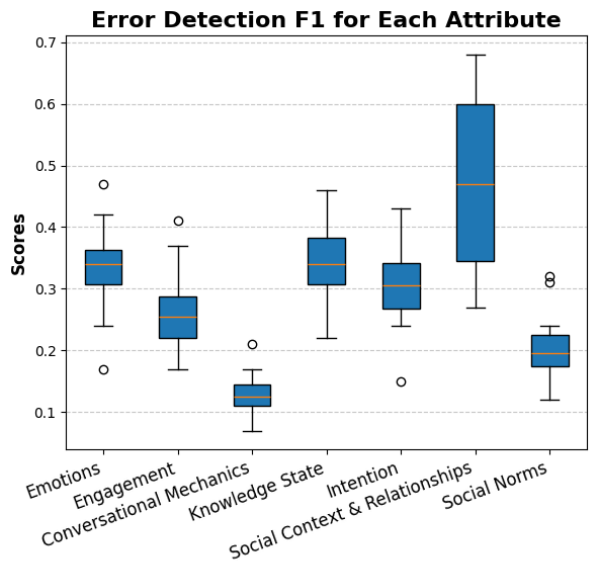}
    \caption{F1 scores on error detection for each attribute. Boxplot is constructed over the results of all 17 models. }
    \label{fig:error_per_attr}
    \end{center}
\end{figure}

We carry out further analysis to identify for which attribute the model is better at detecting errors. This highlights the usefulness of our benchmark, enabling analysis on when LMs fail, for which attribute and why, as our dataset contains aligned comprehensive labels for every annotation. We report the F1 scores of error detection conditioned on the type of attribute the error is related to. While no models perform well on every single attribute, models tend to perform well on the social context and relationship category, whereas they perform poorly on conversational mechanics and social norms. This plot demonstrates strengths and weaknesses of current LMs and identifies pressing areas of research.

\section{Attribute Identification Per Attribute}
\label{app:attr_by_attr}
\begin{figure}[htb]
    \begin{center}
    \vspace{-5mm}
    \includegraphics[width=0.5\linewidth]{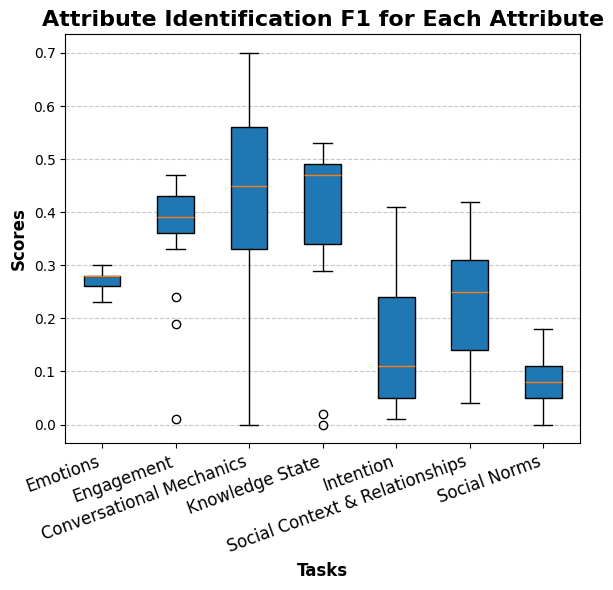}
    \caption{Attribute Identification F1 Per Attribute}
    \label{fig:ident_attr}
    \end{center}
\end{figure}

In Fig. \ref{fig:ident_attr} we carry out further analysis to identify for which attributes model is good at identifying. This analysis showcases, given the true label for error or competency, whether the model is able to identify the related attribute. We find that given the error or competency label, still struggles to identify the correct attribute. More specifically, its performance on social norms, intention, social context and relationships are quite low. We find that certain models perform better than others in specific attributes GPT-4o variants excel in predicting attributes such as emotion, engagement, knowledge state, and intention. In contrast, Llama-3.2 demonstrates strength in identifying social context and relationships, while Gemini shows better  performance in conversational mechanics.

\section{Annotation Procedure}
\label{app:instructions}
\begin{figure}[ht]
    \begin{center}
    \includegraphics[width=0.45\textwidth]{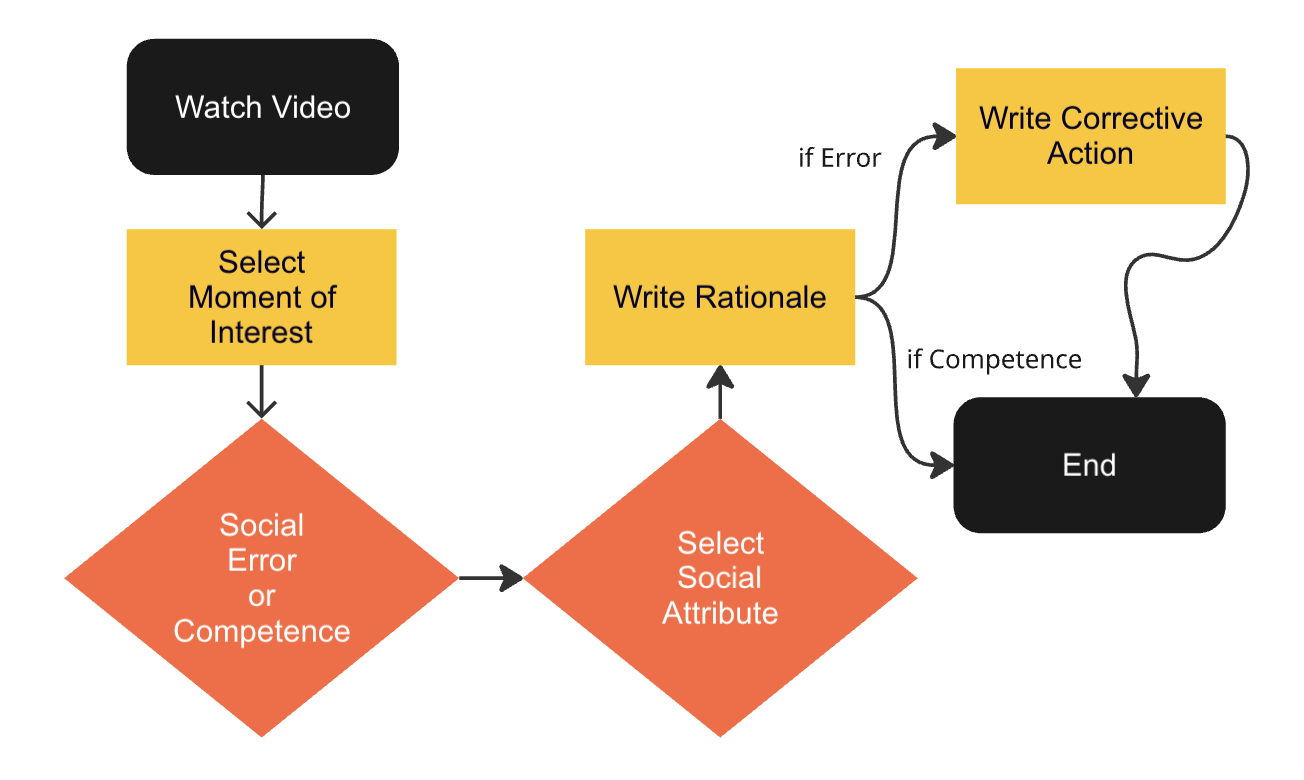}
    \caption{Annotation Procedure: Annotators watch the video, select moment of interest, then label for  error or competence, then they select the social attribute and are asked to write the reason and alternate corrective behavior.}
    \label{fig:anot}
    \end{center}
\end{figure}

You are be given a video of a social robotic agent and you will be asked to annotate the agent's behavior where it exhibits social competence or an Social error. Here we share the definitions:

\textbf{Social Competence:} Social competence is the ability to successfully conduct social interactions, which depends on the awareness and identification of social-emotional cues, the ability to process such cues, and the ability to decide on and express a normative response to these cues.

\textbf{Social Error:} are errors that violate social norms and degrade a user’s perception of a robot’s Social competence, such as interrupting a user at an inappropriate time during a conversation

Simply put, Social competence refers to skillful social and affective behavior that is aligned to the desired and/or normal behaviors expcted by a user, thereby increasing trust, reliability, and overall perceived competence of the agent. Social error refers to a behavior exhibited by a robot that deviates from the desired or normal behaviors expected by a user, thereby degrading the overall perceived competence of the agent. As you may be able to tell, Social competence and error refers to behaviors that deviate or are  aligned to the user's expectations. This specifically involves first recognizing, then responding appropriately to social and affective contexts. Here we share specific social attributes and definitions.

\begin{itemize}
    \item \textbf{Emotions:} The ability to identify and interpret emotional expressions in oneself and others, allowing for empathetic responses and social awareness, e.g. recognizing that someone crying might mean they're sad \cite{golan2006reading}.
    \item \textbf{Engagement:} The skill to observe and assess levels of participation and involvement in social interactions, including cues that indicate interest or disinterest, e.g. continuing to tell a story when a listener is engaged \cite{davis1980interpersonal}. 
    \item \textbf{Conversational Mechanics:} Understanding the structure and flow of conversations, including turn-taking, interruptions, and cues for when to speak or listen, e.g waiting for another person to finish speaking before speaking \cite{fusaroli2016investigating}. 
    \item \textbf{Knowledge State:} The ability to assess what others know or believe, as well as being aware of one's own knowledge in social situations, e.g. user talks about their dog, remembering that the user has a dog \cite{baron1999new}. 
    \item \textbf{User Intention:} The capacity to infer the goals or purposes behind the actions and words of others, facilitating better responses in social interactions, e.g. when the user says ``I'll be right back", indicates that the user will vacate and then return  \cite{dziobek2006introducing}.
    \item \textbf{Social Context and Relationships:} The ability to identify and understand the dynamics of social relationships and the context in which they occur, influencing behavior and expectations, e.g. knowing how to act in front of a close friend vs colleague at work \cite{baron1999new}. 
    \item  \textbf{Social Norms and Routines:} The skill to identify accepted behaviors and attitudes within a social group, as well as recognizing negative or harmful interactions that violate these norms e.g. understanding that waving hands in the beginning of the interaction is a sign of a greeting \cite{thoits2004emotion}.
\end{itemize}

Then the annotators utilize the following annotation tool. 

\begin{figure*}[t]
    \begin{center}
    \includegraphics[width=1\textwidth]{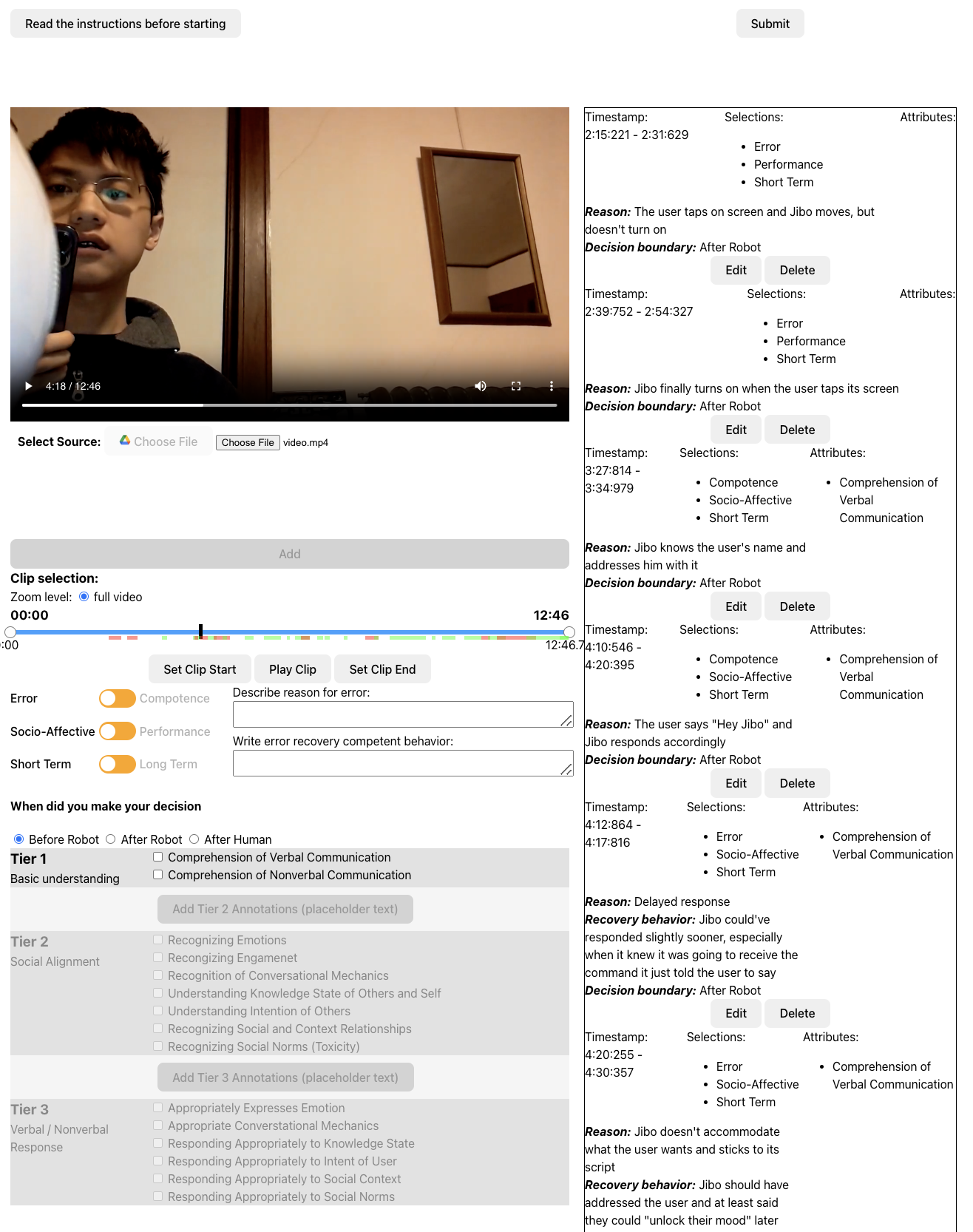}
    \caption{Participant Consented to share images for publications. A screenshot of our annotation tool. Our tool enables the viewing of the video interaction, marking the moments of interest via sliders (after marking errors are in red, competencies are in green below the sliders, annotating for specific attributes). Typos were post-processed. }
    \label{fig:start}
    \end{center}
    \vspace{-4mm}
\end{figure*}
\begin{figure*}[t]
    \begin{center}
    \includegraphics[width=1\textwidth]{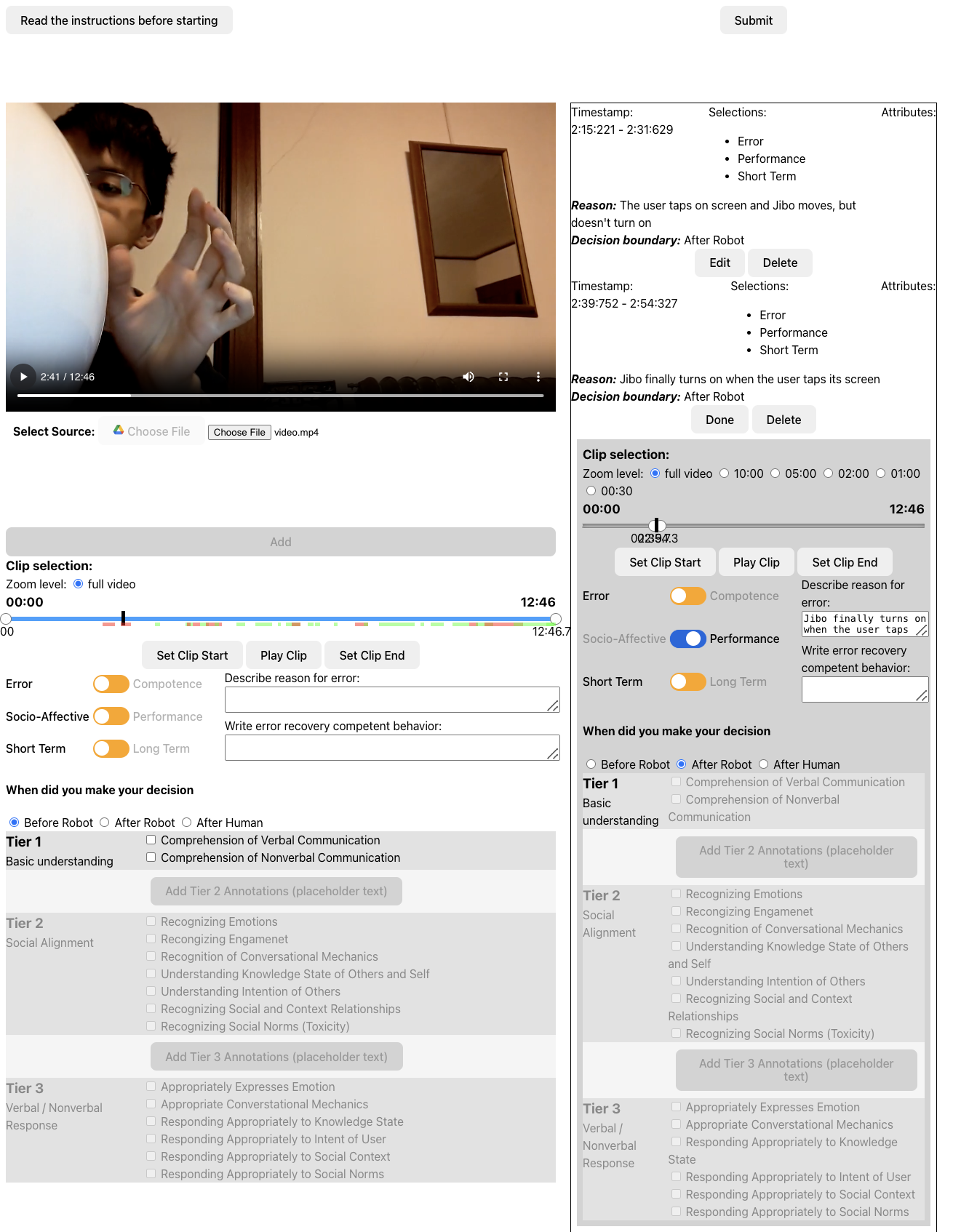}
    \caption{Participant Consented to share images for publications. A screenshot of our internal annotation tool in the edit phase. Our tool flexibly allows the annotator to edit their previous annotations and look back at them. Typos were post-processed.}
    \label{fig:edit}
    \end{center}
    \vspace{-4mm}
\end{figure*}

\textbf{Annotator Consistency} This protocol aligns with established best practices for achieving reliable subjective annotations \cite{artstein2008inter}. We further elaborate below. Each video was annotated independently by two trained annotators, who were free to identify segment boundaries for social errors and competencies. This flexible segmentation approach ensured that annotations captured the natural flow of interaction but introduced potential variability. Across the dataset, 85.1\% of annotations overlapped, meaning that both annotators independently marked the same temporal segment. To ensure consistency, annotators underwent joint training with shared definitions and guidelines. Disagreements were initially addressed through collaborative review sessions on a held-out set, allowing annotators to align their interpretation of ambiguous or edge cases. After this calibration phase, annotators continued to annotate independently. For all non-overlapping or ambiguous annotations, the annotators reviewed each other's work and explicitly marked whether they agreed or disagreed with the alternate annotation. The remaining 8.7\% of annotations were marked as persistent disagreements, reflecting segments where subjective interpretation of social behavior could not be reconciled. These were retained to preserve the richness and variability inherent to real-world human–robot interaction. This protocol balances rigor with the acknowledgment that social reasoning is inherently subjective. We view the residual disagreement not as noise, but as a valuable reflection of human subjective interpretations, we believe that it could also pose a meaningful challenge for AI models.

\textbf{Testing Instructions with Independent Annotators} We tested whether the guidelines would generalize to someone who had not participated in the initial discussions. A third annotator joined later in the process and was provided only the written instructions and example videos (without exposure to the earlier calibration sessions). When comparing this annotator’s labels against the existing annotations, we observed a high agreement of 0.928 across the two original annotators’ annotations, suggesting that the guidelines are clear and reproducible beyond the initial annotator pool.

\section{Anonymization Robustness Experiment}
\label{app:anon_exp}

\begin{figure*}[t]
    \begin{center}
    \includegraphics[width=1\textwidth]{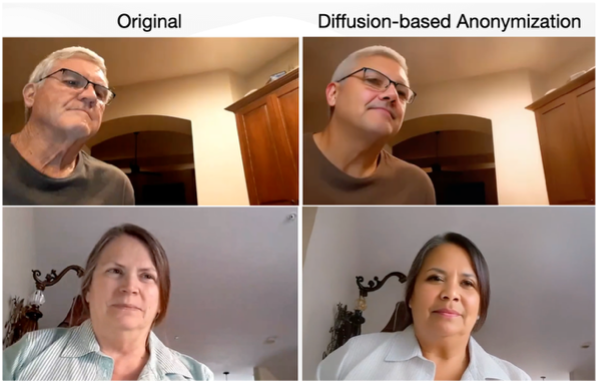}
    \caption{Original face (left) transformed into a fully synthetic version (right), preserving key social while ensuring privacy for responsible large-scale data release.}
    \label{fig:edit}
    \end{center}
    \vspace{-4mm}
\end{figure*}
To explore the impact of anonymization artifacts, particularly from FRESCO, we conducted a controlled experiment comparing vision-language model (VLM) performance across four video conditions:

\begin{itemize}
    \item \textbf{Raw}: Unaltered, raw video
    \item \textbf{Diffusion}: FRESCO, Face replacement via text-guided diffusion (Yang, 2024)
    \item \textbf{Deepfake}: MobileFaceSwap replaces the face with random face images (Xu, 2022)
    \item \textbf{Cartoon}: VToonify stylized, animated face rendering (e.g., Disney style) (Yang, 2022)
\end{itemize}

We evaluated three VLMs---GPT-4o, Gemma-3, and LLaVA-Next---on a 3-way classification task (\textit{Social Competence, Social Error, None}) using 10 balanced runs (15 samples each). Table~\ref{tab:anonymization_results} reports the average F1 and accuracy scores across conditions.

\begin{table}[h]
\centering
\caption{Average F1 and accuracy scores (mean $\pm$ std) across anonymization conditions.}
\label{tab:anonymization_results}
\footnotesize
\setlength{\tabcolsep}{4pt}
\renewcommand{\arraystretch}{1.1}

\resizebox{\columnwidth}{!}{%
\begin{tabular}{lcccc}
\toprule
\textbf{Model} & \textbf{Raw} & \textbf{Diffusion (FRESCO)} & \textbf{Deepfake} & \textbf{Cartoon} \\
\midrule
GPT-4o      & $0.462 \pm 0.083$ & $0.414 \pm 0.096$ & $0.513 \pm 0.058$ & $0.500 \pm 0.063$ \\
Gemma-3     & $0.194 \pm 0.050$ & $0.144 \pm 0.010$ & $0.196 \pm 0.045$ & $0.198 \pm 0.044$ \\
LLaVA-Next  & $0.266 \pm 0.104$ & $0.309 \pm 0.119$ & $0.298 \pm 0.120$ & $0.260 \pm 0.075$ \\
\bottomrule
\end{tabular}%
}
\end{table}

For GPT-4o, we observed a slight performance drop when using FRESCO compared to Raw and Deepfake videos. Interestingly, for LLaVA-Next, FRESCO slightly improved performance, possibly due to denoising or abstraction effects that helped the model focus on salient social cues.

Overall, Cartoon and Deepfake performed comparably or slightly better than Raw across models. Importantly, statistical analysis revealed no significant difference ($p > 0.05$) between Raw and any anonymized variant, including FRESCO.

These results suggest that even with visible visual artifacts, FRESCO does not significantly degrade model performance. We thank the reviewer for this suggestion and will include this experiment in the final version of the paper.
\color{black}

\section{Limitations}
\label{app:limit}
There are several limitations to our study that warrant discussion. First, the scope of social attributes we focus on—while grounded in existing frameworks—is not exhaustive. Social behavior is highly multifaceted, and other relevant competencies or error types may fall outside the seven social attributes we annotated. This limited coverage may reduce the generalizability of our findings to broader or less structured social scenarios.

Second, our annotation process relies on human judgment, which introduces potential subjectivity. While we implemented annotation guidelines and validation checks to improve consistency (see Appendix G), edge cases and disagreements (notably ~8.7\%) suggest that some annotations reflect annotator bias or ambiguity in interpretation. These effects may be compounded in complex, multi-label settings where social attributes co-occur, as shown in our attribute co-occurrence analysis (Appendix B).

Third, our dataset and experiments are geographically and culturally constrained, as all interactions were recorded in the United States. Social norms, conversational practices, and interpretations of robot behavior can vary significantly across cultures, limiting the cross-cultural robustness of both the benchmark and the model evaluations. Similarly, while we included a range of demographic backgrounds, some underrepresented subpopulations (e.g., elderly users, neurodivergent individuals) are insufficiently covered, which may impact the applicability of our findings in more diverse settings.

Fourth, there are strong modeling assumptions underlying many of the foundational models evaluated—such as independence between input modalities (in some architectures), idealized training corpora, and noise-free text or video representations. These assumptions may not hold in real-world settings involving ambiguous, noisy, or unstructured interaction data. Furthermore, our benchmark evaluates models at 1 Hz video sampling, which may miss subtle but socially relevant temporal dynamics.

Fifth, while we evaluate 17 modern LLMs and VLMs, each model is typically run with a single set of configurations and prompts, and we do not report variance across different random seeds or fine-tuning strategies. This limits the strength of claims about model generalization or robustness, especially given that some model performance differences may be attributable to configuration rather than inherent capability.

Sixth, due to the scarcity of large-scale real-world datasets in HRI, we prioritized collecting as much high-quality interaction data as possible across varied contexts. It is correct that the majority of videos in SHREC primarily focus on the human participant, with the robot partially visible in many recordings. We acknowledge that this limits the dataset’s ability to support systematic evaluation of the robot’s full-body non-verbal behaviors or control policies. We will explicitly note this as a limitation in the final version and we also believe this highlights an important direction for future dataset collection: pairing human-centric recordings with synchronized robot-camera views or external wide-angle shots to better capture embodied robot behavior for modeling.

Our task formulations inevitably introduce sources of ambiguity that may affect both model and human performance. First, the Interaction Progression prompt relies on the notions of “pre-condition” and “post-condition” to indicate the likely motivation preceding an agent’s action and the plausible continuation following a user’s action. While we adopted these terms from prior NLP literature, they differ from their meaning in planning theory and may not perfectly capture the response–adjacency relation we intend. This choice of terminology and the conflated visual presentation of the two prompts could lead to confusion for both annotators and models. Moreover, because conversational responses often admit multiple reasonable antecedents or continuations (e.g., “yes” or “ok” may be appropriate in a wide range of contexts), the task may not have a uniquely correct answer in all cases. This introduces unavoidable variability in human judgments and may limit the interpretability of accuracy as a strict measure of correctness.

Furthermore, the construction of distractor items presents challenges. For Interaction Progression, multiple distractors may remain partially acceptable depending on how broadly one interprets conversational adjacency. For the rationale and correction tasks, our strategy of selecting distractors with different social-attribute annotations than the ground-truth response ensures surface-level distinctness, but may also reduce task difficulty. In particular, some incorrect rationales may differ sharply from the correct one, making the choice easier than intended. Conversely, there are cases where an action that remedies one type of social error could also plausibly remedy another, raising the possibility that more than one option could be considered correct. Thus, while our distractor selection method enforces consistency, it cannot guarantee the nuanced ambiguity of real-world social interactions.

Lastly, our current benchmark does not explicitly address fairness, bias, or privacy concerns beyond anonymization. While we use FRESCO for high-fidelity face stylization to preserve social signals and protect identity, potential demographic biases in model outputs (e.g., different error rates across user identities) remain unexplored. Future work should assess whether foundational models trained on SHREC exhibit biases in behavior interpretation or correction, particularly across identity markers such as gender, race, or age.

\section{Broader Impacts}
\label{app:broad}

Our study addresses both the positive and negative societal impacts of our work, namely in how the SHREC dataset and benchmarks can pave the way for developing more socially adept robots that can be used to assist various populations in need. There is also a brief discussion of how possible bias may negatively impact perceptions of social behavior that the robot may be interpreting.

The creation of the SHREC dataset and its associated benchmarks stand to significantly advance the development of socially intelligent AI agents through grounding evaluation and training in actual human-robot interactions. By enabling a more fine-grained analysis and evaluation of social competencies and errors in real-world interactions, our study contributes a foundational resource for improving human-robot communication and interaction quality. Applications of this study could enhance assistive robotics, eldercare, education, and mental health support, where empathetic and socially appropriate behavior from robots can meaningfully improve user experience and well-being.

The societal deployment of socially adept AI agents does carry some notable risks. The improved social fluency in robots could lead to an overreliance on AI systems, particularly in vulnerable populations such as children or the elderly. This nuanced mimicry of human-like empathy by non-sentient machines may blur ethical lines in perceived agency and accountability. Additionally, there is a risk that biases in the interaction annotations may reinforce normative assumptions about social behavior, leading to potential exclusion or misinterpretation of diverse cultural or neurodivergent communication styles. These concerns emphasize the importance of ethical safeguards, continued human oversight, and more inclusive design practices as research in this area progresses.

\paragraph{Environmental Impact}While the computational requirements for this work are modest compared to many large-scale AI projects, they still contribute to the overall carbon footprint of research. Based on an estimate of approximately 480 GPU-hours, we calculate an energy consumption of roughly 192 kWh (assuming a 0.4 kW power draw per GPU). Using a global average carbon intensity of 0.4 kg $CO_2$e/kWh, this corresponds to approximately 77 kg $CO_2$e, comparable to the emissions from driving an average passenger vehicle approximately 190 miles. We acknowledge that actual emissions will vary by GPU hardware, utilization, and local energy mix, and we encourage future work to report detailed compute and energy use. As part of our ongoing work, we are exploring strategies to minimize training runs, increase hardware efficiency, and utilize lower-carbon energy sources where possible.

\paragraph{Safety and Security \& Deception and Harassment} All three source studies that comprise SHREC (Wellness-Dorm, Wellness-Home, and Empathic++) were conducted under approved Institutional Review Board (IRB) protocols at our institutions, which explicitly address these issues. As part of the consent and briefing process, participants were clearly informed that the robot was not sentient, that its social behaviors were scripted or AI-generated, and that its role was to support a research study, not to provide professional, medical, or psychological advice.For example, consistent with ethical research practices,in the Wellness studies, we explicitly informed participants that the robot was not designed to assist with mental health emergencies and had limited perceptual capabilities (e.g., “Robot ears are different from human ears and I (Jibo) might have trouble understanding what you say”). Each robot station displayed a sticker with the National Suicide Prevention Lifeline hotline information. Pre-screening included the PHQ-9 depression questionnaire, and our protocol specified that anyone scoring above 20 (max score 27) would be excluded from participation; in practice, no participants met this exclusion threshold. These measures ensured that participants were aware of the robot’s limitations and that appropriate safeguards were in place for any mental health concerns. Similarly, the onboarding protocol for the Empathic++ study explicitly informed participants that the robot lacked sentience and that its utterances were produced by OpenAI’s GPT model.

\paragraph{Data Quality and Representativeness \& Discrimination, Bias, and Fairness} Our current dataset, drawn from three prior IRB-approved studies (Wellness-Dorm, Wellness-Home, and Empathic), reflects the participant pools of those original deployments. In the Wellness \cite{jeong2020robotic, jeong2020robotic} studies (N=70), participants represented multiple racial and ethnic backgrounds: White (62.8\%), Asian/Pacific Islander (28.5\%), Black or African American (2.85\%), Hispanic/Latino (2.85\%), Native American (1.42\%), and multi-racial (1.42\%), with a mean age of 46 years and a standard deviation of 23 (Range: 18-83 years old) and consisted of consisted of 65.7\% females, 28.6\% males, and 5.7\% others.. The Empathic (N=46) \cite{shen2024empathicstories++} study participants ranged from 20-75 years old with a mean age of 36 and a standard deviation of 14.45, and contains 38.9\% males and 61.1\% females . While these samples provide rich, real-world interaction data, they are U.S.-centric and underrepresent certain age groups, particularly older adult populations. We will acknowledge this in the paper’s Limitations section, but also stress that this work represents one of the largest, most richly annotated real-world social robot interaction datasets to date. Importantly, expanding diversity is an active and ongoing effort. Firstly, we have conducted prior studies with both older adults and children, and we plan to gradually integrate these interactions into the dataset to broaden its demographic coverage over time \cite{park2017growing,ostrowski2022mixed}. Secondly, our group leads the Jibo Community Social Robot Research Platform @Scale \cite{park2024jibo} initiative, we have converted previously commercialized Jibo robots into a shared, cloud-connected, community research platform. This infrastructure is designed to enable social robot living labs—multi-site, long-term deployments in diverse settings such as homes, schools, community centers, and senior living communities. By partnering with a broad network of researchers, including those embedded in underserved communities, we are already laying the groundwork for future SHREC expansions with broader cultural, geographic, and age diversity, ensuring the benchmark grows even more representative over time.

\begin{figure*}[t]
    \begin{center}
    \includegraphics[width=1\textwidth]{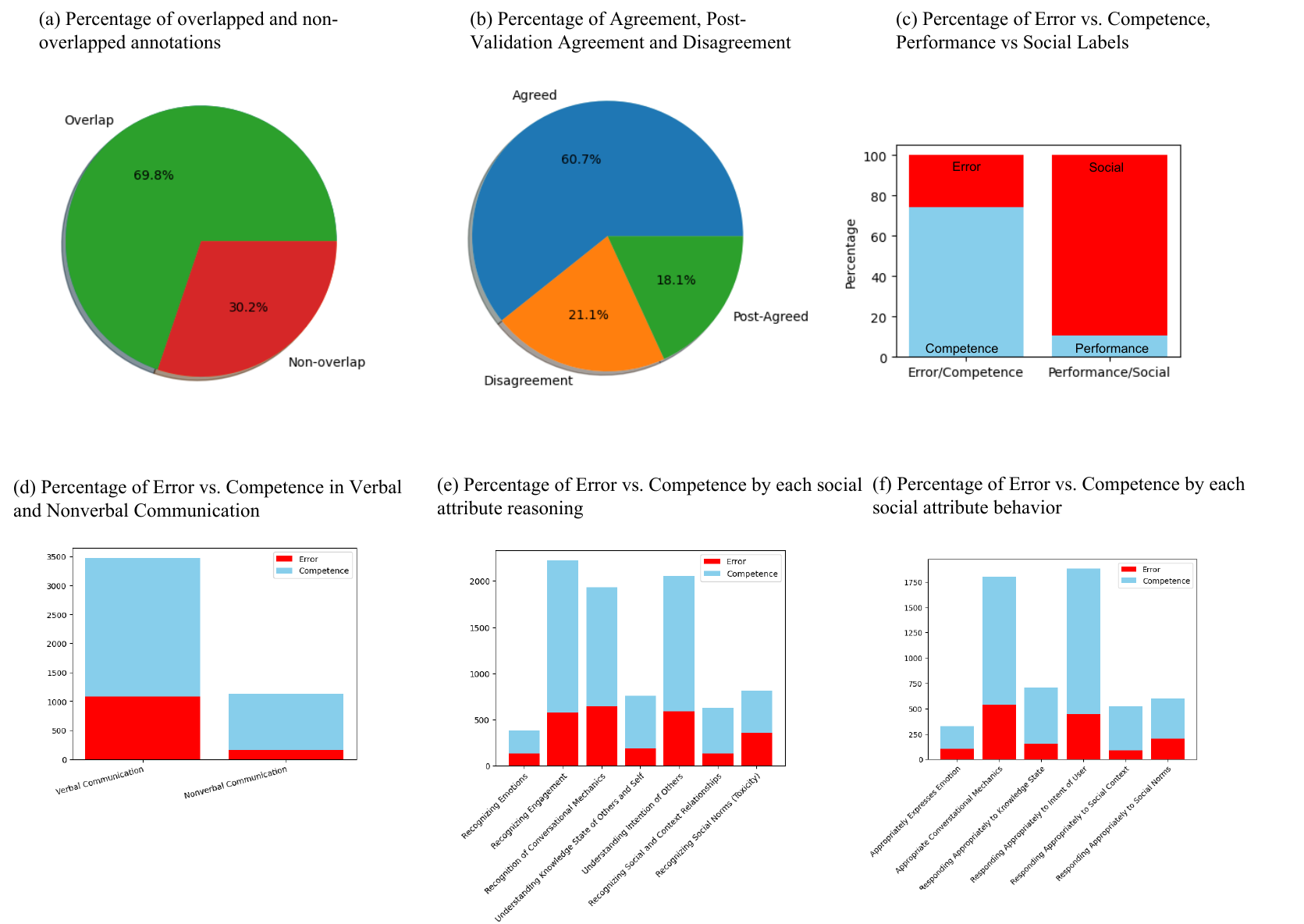}
    \caption{Wellness \cite{jeong2023deploying} Dataset Statistics: We find that 69.8\% of the dataset consists of overlapping annotations. Amongst the overlapping samples as shown in Figure B, we find an 78.1\% overall agreement, where annotators agree on the error/competency and social/competency labels. In Figure C, again, we showcase the percentage of errors and competencies on the left and whether if they were related to social or performance. We find the majority being competencies relating to social dimensions. In the bottom row, we showcase plots regarding whether the error or competencies manifested in the perception, or the reasoning, or the behavior. In figure (d), we find that majority of the annotations marked by annotators belong in the verbal communication category. In figure (e) and (d), we find that most annotations belong in understanding or responding to (1) recognizing engagement, (2) conversational mechanics, (3) intent. . If we consider the competencies and errors separately, we find that annotators marked the most number of errors for conversational mechanics, intent and knowledge state and most number of competencies for engagement and social context. }
    \label{fig:wellness_stats}
    \end{center}
\end{figure*}
\begin{figure*}[t]
    \begin{center}
    \includegraphics[width=1\textwidth]{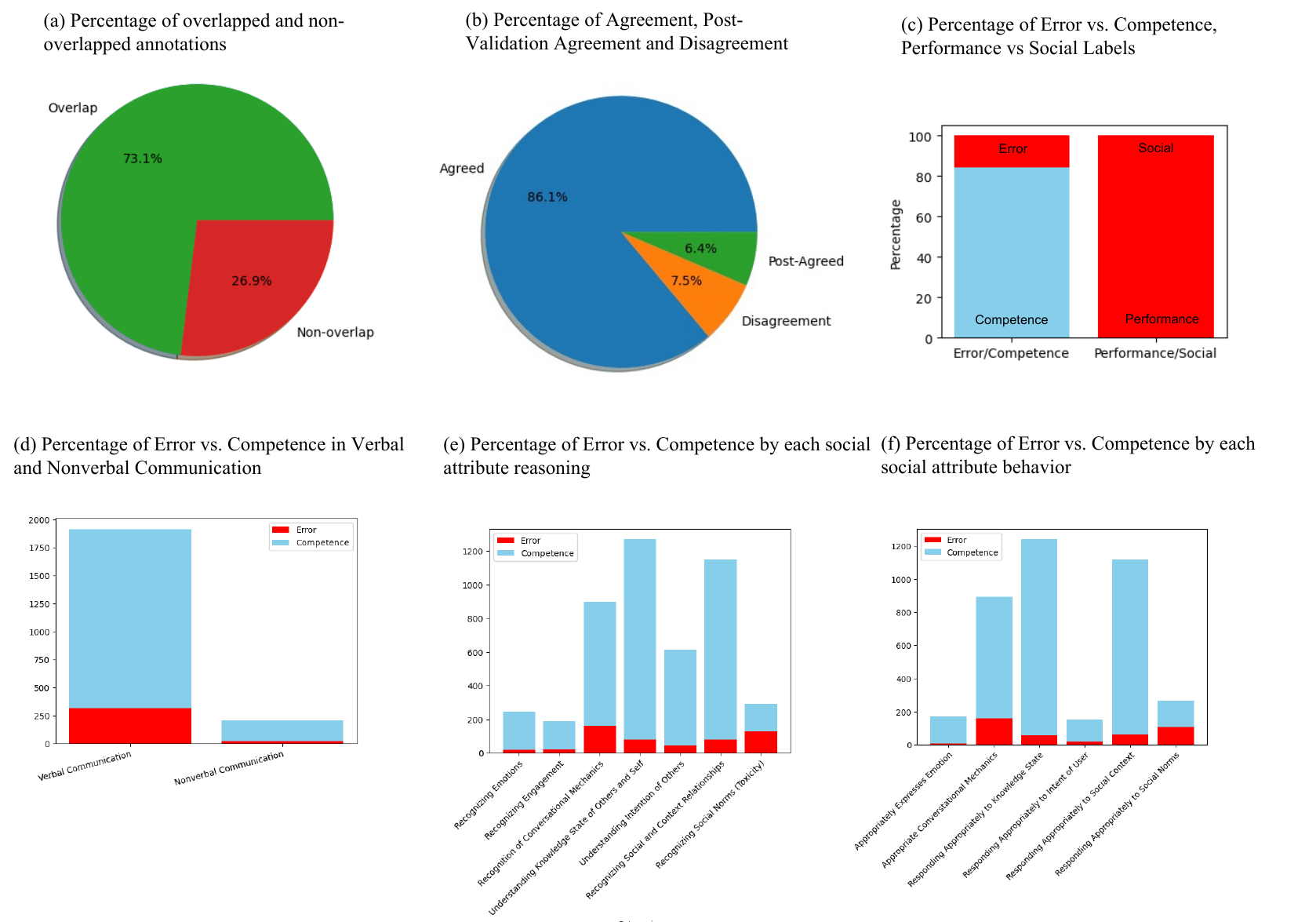}
    \caption{Empathic \cite{shen2024empathicstories++} Dataset  Statistics:  We find that 73.1\% of the dataset consists of overlapping annotations, where two annotators marked the sample . We refer the reader to Appendix REFER for the algorithm used to calculate overlaps. Amongst the overlapping samples as shown in Figure B, we find an 92.5\% overall agreement, where annotators agree on the error/competency and social/competency labels. A random agreement would have been 25\%. We go through a two-step procedure of this process, where in the first phase, the annotator simply annotates the video themselves. In the second step, the annotator looks at the other annotator’s annotation and agrees whether or not this could be a possible interpretation. In Figure C, we showcase the percentage of errors and competencies on the left and whether if they were related to social or performance. We find the majority being competencies relating to social dimensions. In the bottom row, we showcase plots regarding whether the error or competencies manifested in the perception, or the reasoning, or the behavior. In figure (d), we find that majority of the annotations marked by annotators belong in the verbal communication category. In figure (e) and (d), we find that most annotations belong in understanding or responding to (1) knowledge state, (2) social relationships, (3) conversational mechanics. If we consider the competencies and errors separately, we find that annotators marked the most number of errors for conversational mechanics, and toxicity and most number of competencies for knowledge state and social context. 
}
    \label{fig:empathic_stats}
    \end{center}
\end{figure*}

\section{Assets License}
\label{app:assets}

We benchmarked 17 large language and vision-language models, including both open-source and proprietary systems, on the SHREC benchmark. We also constructed our dataset using three prior human--robot interaction studies. Below we list each asset along with its creator and license or usage information, where available.

\subsection*{Models}

\begin{itemize}
    \item \textbf{GPT-4o}, \textbf{GPT-4o-mini},  \textbf{o1}  
    \begin{itemize}
        \item Creator: OpenAI  
        \item License: Accessed via API under \href{https://openai.com/policies}{OpenAI Terms of Use}
    \end{itemize}
    
    \item \textbf{Gemini 1.5}, \textbf{Gemini 2.0}  
    \begin{itemize}
        \item Creator: Google DeepMind  
        \item License: Accessed via API under \href{https://ai.google.dev/terms}{Google AI Usage Terms}
    \end{itemize}
    
    \item \textbf{DeepSeek-R1-Distill-Qwen-32B}  
    \begin{itemize}
        \item Creator: DeepSeek AI  
        \item License: MIT License; \href{https://huggingface.co/deepseek-ai/DeepSeek-R1-Distill-Qwen-32B}{Hugging Face link}
    \end{itemize}
    
    \item \textbf{LLaMA-3.2}  
    \begin{itemize}
        \item Creator: Meta AI  
        \item License: Meta Llama 3 Community License; \href{https://github.com/meta-llama/llama-models/blob/main/LICENSE}{License}
    \end{itemize}
    
    \item \textbf{LLaMA-3.2-11B-Vision-Instruct}  
    \begin{itemize}
        \item Creator: Meta AI  
        \item License: Same as above
    \end{itemize}
    
    \item \textbf{LLaVA-Next-LLaMA3}  
    \begin{itemize}
        \item Creator: LLaVA Team / Meta AI  
        \item License: Built on LLaMA-3.2; inherits Meta's license
    \end{itemize}
    
    \item \textbf{InternVL2-8B}  
    \begin{itemize}
        \item Creator: OpenGVLab  
        \item License: MIT License; \href{https://huggingface.co/OpenGVLab/InternVL2-8B}{Hugging Face link}
    \end{itemize}
    
    \item \textbf{MiniCPM-V 2.6}  
    \begin{itemize}
        \item Creator: OpenBMB  
        \item License: Apache 2.0 (code); MiniCPM Model License (weights); \href{https://huggingface.co/openbmb/MiniCPM-V-2_6}{Hugging Face link}
    \end{itemize}
    
    \item \textbf{PaliGemma}  
    \begin{itemize}
        \item Creator: Google / Gemma  
        \item License: Apache 2.0 (code), CC-BY 4.0 (content); \href{https://ai.google.dev/gemma/docs/paligemma}{Docs}
    \end{itemize}
\end{itemize}

\subsection*{Datasets}

The SHREC dataset comprises real-world social interactions collected across three previously unpublished studies, now shared under a new IRB protocol. All data releases follow institutional guidelines for responsible dissemination.

\begin{itemize}
    \item \textbf{Empathic++}~\cite{shen2024empathicstories++}: A ChatGPT-powered empathic social robot facilitated emotionally meaningful conversations using narrative therapy techniques to promote connection and belonging.
    
    \item \textbf{Wellness-Dorm}~\cite{jeong2020robotic}: A socially assistive robot served as a positive psychology coach for college students in dormitories, delivering interventions on gratitude, strengths reflection, and goal-setting.
    
    \item \textbf{Wellness-Home}~\cite{jeong2023robotic}: Robots were deployed in participants’ homes under three conditions: assistant, coach, and companion, each offering varying degrees of social and functional support.
\end{itemize}

All datasets were anonymized using \texttt{FRESCO}, a diffusion-based video anonymization framework (\href{https://github.com/martiniani-lab/FReSCo}{MIT License}), and manually filtered to remove personally identifiable information. The SHREC dataset is released via gated access on \href{https://huggingface.co/datasets/MIT-personal-robots/shrec}{Hugging Face}, with use governed by institutional IRB approvals.

\section{LLM Usage}

We used large language models (LLMs), specifically OpenAI’s ChatGPT, as a writing and editing aid during the preparation of this manuscript. The LLM was employed to polish wording, improve clarity and flow, and rephrase certain sections for readability. The final text was reviewed and edited by the authors to ensure accuracy and originality.

\end{document}